\documentclass[twocolumn,twocolappendix]{aastex631}

\usepackage{graphicx}

\newcommand{\mum}{\ifmmode{\rm \mu m}\else{$\mu$m}\fi}

\shorttitle{Mid-infrared spectra of LMC carbon stars}
\shortauthors{Sloan et al.}

\begin{document}

\title{Mid-infrared JWST spectra of carbon stars in the Large Magellanic Cloud}

\correspondingauthor{G.~C. Sloan}
\email{gcsloan@stsci.edu}

\author[0000-0003-4520-1044]{G.~C.\ Sloan}
\affiliation{Space Telescope Science Institute, 3700 San Martin Drive,
             Baltimore, MD 21218, USA}
\affiliation{Department of Physics and Astronomy, University of North
             Carolina, Chapel Hill, NC 27599-3255, USA}
\author[0000-0001-9848-5410]{B.\ Aringer}
\affiliation{Theoretical Astrophysics, Department of Physics and Astronomy,
             Uppsala University, Box 516, 751 20 Uppsala, Sweden}
\affiliation{Department of Astrophysics, University of Vienna,
             T\"{u}rkenschanzstra{\ss}e 17, 1180 Wien, Austria}
\author[0000-0002-2626-7155]{Kathleen  E.\ Kraemer}
\affiliation{Institute for Scientific Research, Boston College, 140
             Commonwealth Avenue, Chestnut Hill, MA 02467, USA} 
\author[0000-0002-2666-9234]{J.\ Cami}
\affiliation{Department of Physics and Astronomy, The University of Western
             Ontario, London, ON N6A 3K7, Canada}
\affiliation{Institute for Earth and Space Exploration, The
             University of Western Ontario, London, ON N6A 3K7, Canada}
\author[0000-0002-1614-8195]{K.\ Eriksson}
\affiliation{Theoretical Astrophysics, Department of Physics and Astronomy,
             Uppsala University, Box 516, 751 20 Uppsala, Sweden}
\author[0000-0003-2356-643X]{S.\ H\"{o}fner}
\affiliation{Theoretical Astrophysics, Department of Physics and Astronomy,
             Uppsala University, Box 516, 751 20 Uppsala, Sweden}
\author[0000-0003-1689-9201]{K.\ Justtanont}
\affiliation{Chalmers University of Technology, Dept. Space, Earth and
             Environment, Onsala Space Observatory, 439 92 Onsala, Sweden}
\author[0000-0002-1335-5623]{E.\ Lagadec}
\affiliation{Universit\'{e} C\^{o}te d'Azur, Observatoire de la C\^{o}te
             d'Azur, CNRS, Laboratoire Lagrange, Bd de l'Observatoire,
             CS 34229, 06304 Nice Cedex 4, France}
\author[0000-0002-9137-0773]{Paola Marigo}
\affiliation{Department of Physics and Astronomy G. Galilei, University of 
             Padova, Vicolo dell’Osservatorio 3, I-35122 Padova, Italy}
\affiliation{Deceased}
\author[0000-0002-5529-5593]{M.\ Matsuura}
\affiliation{Cardiff Hub for Astrophysical Research and Technology (CHART),
             School of Physics and Astronomy, Cardiff University, The Parade,
             Cardiff CF24 3AA, UK}
\author[0000-0003-0356-0655]{I.\ McDonald}
\affiliation{Jodrell Bank Centre for Astrophysics, The University of 
             Manchester, Manchester, M13 9PL, UK}
\author[0000-0003-2553-4474]{E.~J.\ Montiel}
\affiliation{U.S.\ Naval Observatory, 3450 Massachusetts Ave.\ NW,
             Washington, DC 20392, USA}
\affiliation{SOFIA-USRA, NASA Ames Research Center, MS 232-12, Moffett Field,
             CA, 94035, USA}
\author[0000-0002-6858-5063]{R.\ Sahai}
\affiliation{Jet Propulsion Laboratory, MS 183-900, 4800 Oak Grove Dr.,
             California Institute of Technology, Pasadena, CA 91109, USA}
\author[0000-0002-3171-5469]{A.~A.\ Zijlstra}
\affiliation{Jodrell Bank Centre for Astrophysics, The University of 
             Manchester, Manchester, M13 9PL, UK}

\begin{abstract}

Mid-infrared spectra from the Medium Resolution Spectrometer on the 
James Webb Space Telescope have revealed the molecular chemistry of 
carbon stars in the Large Magellanic Cloud with better 
resolution and sensitivity than previously possible.  Our sample 
spans a range of dust-production rates and includes three relatively 
dust-free semiregular variables and six dustier Mira variables.  All 
were observed 15--20 yr earlier with the Infrared Spectrograph on 
the Spitzer Space Telescope at lower spectral resolution.  The new 
spectra show that the C$_3$ molecule is responsible for a strong 
absorption band centered at 5.2~\mum.  CS is clearly present in some 
of the sample, especially the stars with less dust.  HCN also appears 
to be present.  Some of the spectra have changed significantly 
between the Spitzer epoch and the MRS observations in 2023 and 
2024, and in most cases these changes can be attributed to the 
stellar pulsation cycle.  One exception is the disappearance of a dust 
emission feature at $\sim$18~\mum\ in one of the Miras.  The new 
spectra reveal a dip centered at $\sim$10~\mum, which could arise 
either from an unknown carrier or from variable molecular emission to 
the red and blue.  The presence of this spectral structure on the 
short-wavelength side of the SiC dust emission feature at 
$\sim$11.3~\mum\ along with the broad C$_2$H$_2$ band centered at 
14~\mum\ raise the possibility that some previously reported 
detections of weak SiC dust emission in other carbon stars may not be 
real.

\end{abstract}

\keywords{carbon stars (199); asymptotic giant branch stars (2100); 
  long period variable stars (935);
  circumstellar dust (236); circumstellar gas (238)}

\vspace{24pt}

\section{Introduction \label{sec:intro}} % Sec. 1

Carbon stars dominate the population of evolved stars on the 
asymptotic giant branch (AGB) in the nearby Large and Small Magellanic 
Clouds \citep[LMC and SMC;][]{bla78, bla80, cio03}.  They also 
dominate the production of dust in these metal-poor dwarf galaxies 
\citep{mat09, mat13, boy12, sri16}.  AGB stars turn carbon-rich when 
they produce carbon via the triple-$\alpha$ process \citep{sal52} and 
dredge enough of it up to their surfaces to push the C/O ratio above 
1.0 \citep[e.g.,][]{ren81}.  These stars are unstable to pulsations in 
their envelopes, which trigger the mass loss that will strip them to 
their cores \citep[e.g.,][]{hab96, mat08, lil16, hof18}.  

\cite{kra19} showed that the amplitude of the pulsations is critical 
to the quantity and chemistry of the dust produced \citep[see 
also][]{mcd19}.  Carbon stars pulsating weakly as semiregular 
variables (SRVs) form small amounts of dust dominated by SiC.  Carbon 
stars pulsating strongly enough to be classified as Mira variables 
produce significantly higher amounts of dust, primarily amorphous 
carbon, which dominates the dust as the stars evolve \citep{mar87}.  
Once a carbon star is forming amorphous carbon, the grains are opaque 
enough to be accelerated to escape velocity by radiation pressure from 
the central star and drag the gas along with it \citep[e.g.,][]{wic66, 
woi06, mat10, bla19, eri23}. 

The dust condenses from molecular gas in the atmosphere, but a clear
understanding of the molecular chemistry of that gas is 
lacking.  \cite{che06} provides a good map of the reaction pathways and 
compounds to be expected.  In a carbon-rich atmosphere, CO consumes 
all of the available oxygen, and molecules such as C$_2$H$_2$, HCN, 
SiS, and CS will form.  All four have been identified in the 
mid-infrared spectra of Galactic carbon stars \citep{aok98, aok99}, 
using the Short Wavelength Spectrometer \citep[SWS;][]{deg96, lee03} 
on board the Infrared Space Observatory \citep[ISO;][]{kes96, kes03}.  
Acetylene (C$_2$H$_2$) may be the primary building block of the 
benzene molecule \citep[C$_6$H$_6$;][]{fre89}, which in turn is the 
building block of graphite, amorphous carbon, and polycyclic aromatic 
hydrocarbons \citep[PAHs; e.g.,][]{all89}.  

The launch of the Spitzer Space Telescope \citep{wer04} made it 
possible to obtain sizable mid-infrared spectroscopic samples of 
carbon stars in nearby galaxies in the Local Group, which enabled the 
study of how metallicity affects the photospheric chemistry and the 
production of dust in carbon stars.  Using the Infrared Spectrograph 
(IRS) on Spitzer \citep{hou04}, several studies obtained spectra of a 
total of 144 carbon stars in the LMC \citep[see][and references 
therein]{slo16}.  The LMC has a metallicity that ranges between 30\% 
and 50\% Solar ($-$0.5 $<$ [Fe/H] $<$ $-$0.3) \citep{pia13, cho16}.  
Despite the difference in metallicity, carbon stars in the LMC and 
even the more metal-poor SMC show no notable differences in the 
quantity of dust produced compared to a sample of Galactic carbon 
stars observed by the SWS on ISO \citep{slo08}.  This trend continues 
to carbon stars in even more metal-poor dwarf spheroidal galaxies in 
the Local Group \citep{slo12}.  These results attest to the power of 
helium burning and dredge-ups on the AGB.  Carbon stars are producing 
all the carbon they need to make dust themselves, no matter their 
initial metallicity.  Thus, they are likely contributors to the dust 
in galaxies even in the high-redshift Universe (as discussed in 
Section~\ref{s.context}).

While metallicity does not affect the dust-production rates in carbon 
stars strongly, the molecular component is another matter.  In 
Galactic carbon stars, acetylene, HCN, CS, and SiS are all detected in 
the mid-infrared, but in the LMC and SMC, only acetylene has been 
observed \citep{aok98, mat06}.  This difference could arise from the 
lower metallicities of the stars, but the IRS spectra have a spectral 
resolving power ($R$ $\equiv$$\lambda$/$\Delta$$\lambda$) of 
$\sim$100, which is high enough to study the molecular bands as a 
whole but too low to study the detailed structure within them.  Thus, 
it is possible that low levels of absorption from molecules beside 
acetylene were present but went undetected.

This project follows up on carbon stars in the LMC previously observed 
with the IRS on Spitzer to better understand the nature of the 
molecular absorption in their spectra.  We have observed a sample of 
carbon stars in the LMC with the Medium Resolution Spectrometer 
\citep[MRS;][]{wel15}, an observing mode of the Mid-Infrared 
Instrument \citep[MIRI;][]{wri23} on board the James Webb Space 
Telescope \cite[JWST;][]{gar23}.  

This paper presents an overview of the entire sample and focuses on 
(1) qualitative molecular analysis; (2) temporal spectral variations; 
and (3) their relation to the stellar pulsation cycles.  
Section~\ref{s.sample} introduces the sample of spectroscopic 
targets; Section~\ref{s.obs} describes the observations and data 
reduction; and Section~\ref{s.analysis} presents the results of the 
spectral analysis.  Sections~\ref{s.disc} and \ref{s.sum} discuss 
and summarize the results.  The appendices examine the ancillary data 
used in this paper and supplement the figures from the spectral 
analysis.

\section{Sample \label{s.sample}} % Sec. 2

\begin{deluxetable*}{lrrlccrrcc} % Table 1
\tablecolumns{10}
\tablewidth{0pt}
\tablenum{1}
\tablecaption{The MRS Sample of Carbon Stars in the LMC}
\label{t.sample}
\tablehead{
  \colhead{Target\tablenotemark{a}} & \colhead{R.A.} & \colhead{Decl.} & 
  \colhead{Var.}     & \colhead{Period} & \colhead{Luminosity} & 
  \colhead{[3.6]$-$[4.5]} & \colhead{Integration} & 
  \colhead{IRS} & \colhead{MRS} \\
  \colhead{ } & \colhead{ } & \colhead{ } & \colhead{Type} & 
  \colhead{ } & \colhead{ } & \colhead{ } & \colhead{Time} &
  \colhead{Epoch} & \colhead{Epoch} \\
  \colhead{ } & \multicolumn{2}{c}{(J2000)} & \colhead{ } &
  \colhead{(days)\tablenotemark{b}} &
  \colhead{(L$_{\odot}$)\tablenotemark{c}} & 
  \colhead{(mag)\tablenotemark{d}} & \colhead{(s)} & \colhead{(MJD)} & 
  \colhead{(MJD)}
}
\startdata
J050629     & 76.623347 & $-$68.926346 & SRV  & (218) & 
    3820 $\pm$   3 & $-$0.150 & 5784 & 54484 & 60273 \\
KDM 1691    & 75.903917 & $-$68.560699 & SRV  &  523  &
  13,650 $\pm$ 560 & $-$0.094 & 1340 & 54649 & 60551 \\
WBP 17      & 81.582833 & $-$69.693703 & SRV  &  309  &
    5510 $\pm$ 310 & $-$0.034 & 4224 & 53596 & 60551 \\
WBP 29      & 81.670642 & $-$69.386490 & Mira &  245  &
    5010 $\pm$  50 &    0.076 & 5112 & 53482 & 60274 \\
J051803     & 79.513592 & $-$68.830750 & Mira &  372  &
    3270 $\pm$ 300 &    0.323 & 1560 & 54688 & 60490 \\
J053441     & 83.672587 & $-$69.441864 & Mira &  519  &
    9190 $\pm$ 590 &    0.643 &  540 & 54611 & 60551 \\
MSX LMC 220 & 78.133621 & $-$69.261230 & Mira &  637  &
  15,760 $\pm$ 530 &    0.863 &  540 & 53379 & 60551 \\
MSX LMC 774 & 81.596238 & $-$69.188965 & Mira &  671  &
    8180 $\pm$ 720 &    1.134 &  540 & 53445 & 60551 \\
MSX LMC 736 & 83.278350 & $-$70.509682 & Mira &  690  &
    8210 $\pm$ 620 &    1.375 &  628 & 54611 & 60385 
\enddata
\tablenotetext{a}{The three stars with names beginning with ``J'' have 
the more formal coordinate-based names 2MASS J0506290$-$6855348, 
SSTISAGE1C J051803.23$-$684950.7, and 2MASS J05344142$-$6926307.  This
paper will refer to them by their shorter ``J'' names throughout.}
\tablenotetext{b}{See Appendix~\ref{s.lc}.  Values in parentheses 
are uncertain.}
\tablenotetext{c}{See Appendix~\ref{s.lum}.}
\tablenotetext{d}{From photometry reported by \cite{slo16}.}
\end{deluxetable*}

\begin{figure}[!ht] % Fig. 1
\includegraphics[width=240pt]{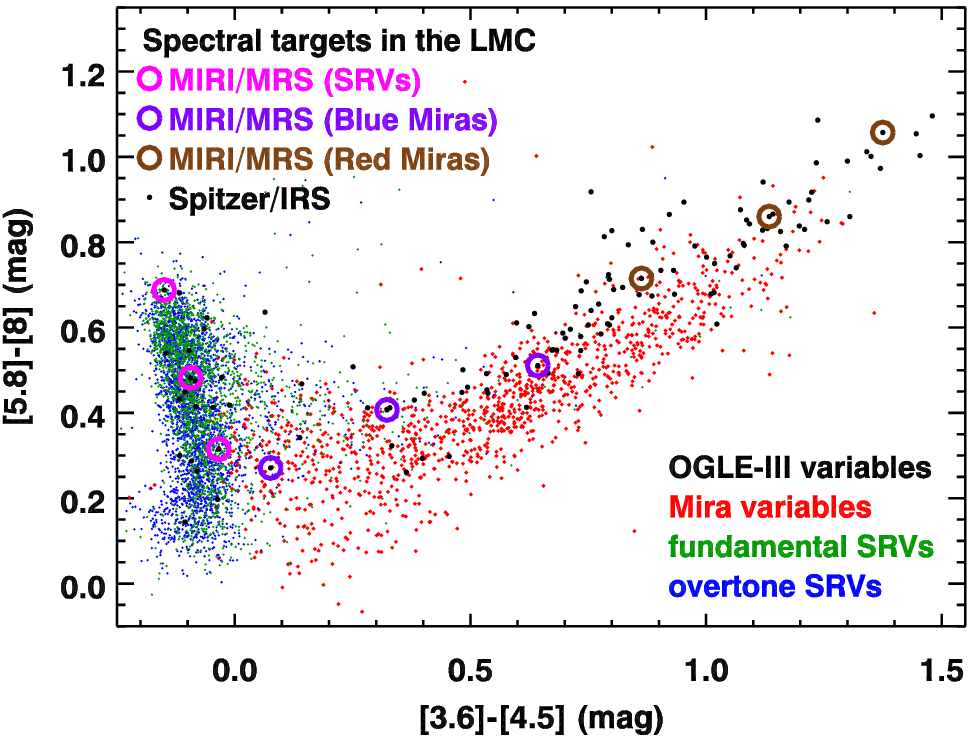}
\caption{The sample of nine MRS targets (open circles) in color-color 
space defined by the four IRAC filters, plotted with data from the
OGLE-III sample of carbon-rich long-period variables in the LMC 
\citep{sos09} and the carbon stars observed by the IRS on Spitzer
\citep{slo16}.  Three of the targets sample the SRV sequence (magenta),
and the other six cover the Mira sequence (purple and brown).
\label{f.sample}}
\end{figure}

JWST program 3010 observed nine carbon stars in the LMC with the MRS
on MIRI in Cycle 2.\footnote{\dataset[DOI: 
10.17909/j0jj-1h55]{https://doi.org/10.17909/j0jj-1h55}}  
Table~\ref{t.sample} provides details on the
properties and observations of each star.  Figure~\ref{f.sample} plots 
the sample in infrared color-color space, along with all 
carbon-rich Mira variables and SRVs from the Optical Gravitational 
Lensing survey of the LMC \citep[OGLE-III;][]{sos09}.  The photometry 
comes from the Infrared Array Camera (IRAC) on Spitzer 
\citep{faz04}, the Wide-field Infrared Survey Explorer 
\citep[WISE;][]{wri10}, and the Near-Earth Object WISE Reactivated 
mission \citep[NEOWISE-R;][]{mai14}.  The data in the 3.4 and 
4.6~\mum\ WISE filters (W1 and W2) have been shifted to the IRAC 3.6 
and 4.5~\mum\ filters using the conversions from \cite{slo16}.

The colors of the carbon stars follow two distinct sequences.  The 
Mira variables, which are undergoing strong fundamental-mode 
pulsations in their atmospheres, lie to the right of the boundary at 
[3.6]$-$[4.5] $\sim$ 0.  The SRVs lie to the left of that boundary, no 
matter their pulsation mode.  \cite{slo15} suggested that the 
range of [5.8]$-$[8] colors in the SRVs could arise from differing 
levels of absorption from C$_3$ at 5~\mum, based on the identification
of C$_3$ in SWS spectra of Galactic carbon stars \citep{loi97, gau04}.  
Deeper C$_3$ absorption would decrease the emission in the 5.8~\mum\ 
filter and redden the [5.8]$-$[8] color.  See Sec.~\ref{s.5um} for new 
observational support of this hypothesis.  In the Mira variables, on 
the other hand, the increasing red colors in all filters result from 
higher optical depths of amorphous carbon dust around them.  

We selected nine targets in Program 3010 to track both the SRV and 
Mira sequences in Figure~\ref{f.sample}, with three stars along the 
SRV sequence and six along the Mira sequence.  We chose targets that 
had been previously observed by the IRS on Spitzer, were roughly 
evenly spaced along the two sequences, and were close to the bar of 
the LMC so that they were observed in six epochs with IRAC on Spitzer.  
The first two IRAC epochs were from the program Surveying the Agents 
of a Galaxy's Evolution (SAGE), which surveyed the entire LMC 
\citep{mei06}.  The other four epochs were obtained in the
SAGE-Var program \citep{rie15}.  

Figure~\ref{f.sample} shows the Spitzer/IRS sample of carbon stars in
the LMC as black dots.  The nine sources also observed by the MRS on
JWST are circled in magenta (for the SRVs) and purple and brown (for
the bluer and redder Miras, respectively).

\begin{figure}[!ht] % Fig. 2
\includegraphics[width=240pt]{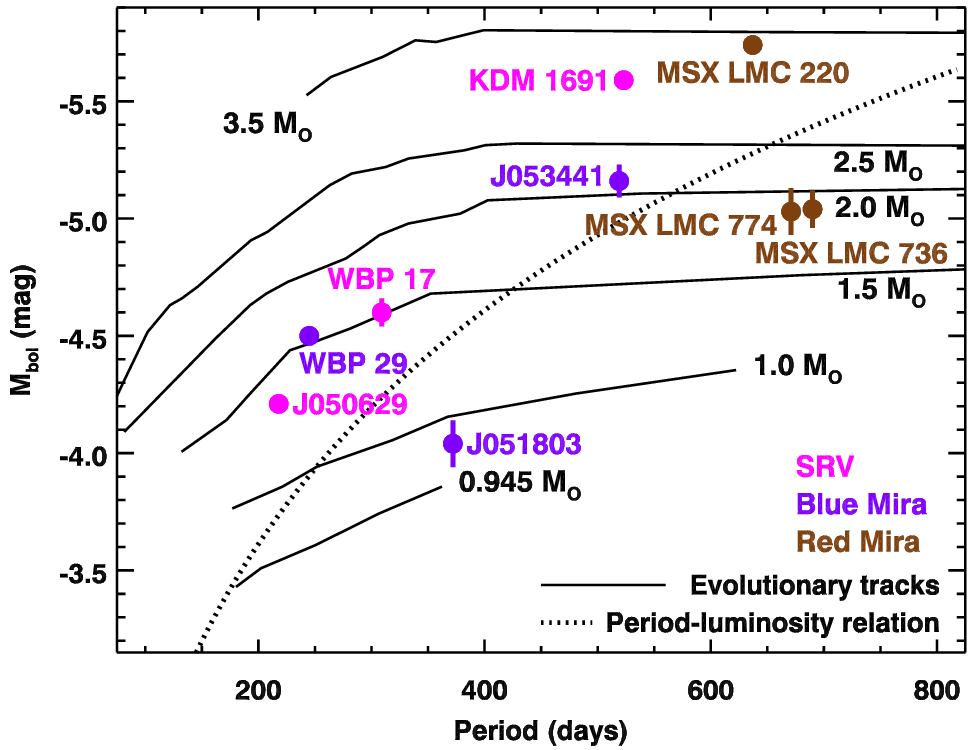}
\caption{The stars in our sample compared to evolutionary tracks by
\cite{vw93} and the period-luminosity relation for carbon stars
\citep{whi09}.  Appendix~\ref{s.lum} describes how the luminosities
were calculated. \label{f.evo}}
\end{figure}

Table~\ref{t.sample} provides the pulsation periods and luminosities 
of the stars in the sample, which were determined as described in
Appendices~\ref{s.lc} and \ref{s.lum}, respectively.  
Figure~\ref{f.evo} compares the properties of the sample to 
evolutionary tracks from \cite{vw93}, which give a rough idea of
the initial mass of the stars.  KDM~1691 and MSX~LMC~220 are 
significantly more luminous than the rest of the sample and thus
more massive and in all likelihood more metal-rich.  J051803 appears
to be the least massive and therefore oldest and most metal-poor star 
in the sample.  The SRVs are all to the left of the period-luminosity
relation determined by \cite{whi09}, which is arguably consistent
with overtone pulsation modes, but the spread in the Miras to either
side of the relation indicates that the relation must be broad and not 
well defined by a single curve.

\section{Observations and data reduction \label{s.obs}} % Sec. 3

Each of the nine stars in the sample was observed in all three grating 
settings in the MRS, giving complete coverage at a spectral resolving
power of $\sim$2000--3000 from 4.9 to 28~\mum.  The 
combination of decreasing emission from the stars and lower 
sensitivity in Channel 4 of the MRS (18--28~\mum\ make the spectra 
less meaningful past a limit of $\sim$19--22~\mum, depending on the 
brightness of the star.  Integration times were set to achieve a 
minimum signal-to-noise ratio (S/N) of $\sim$100 from 5 to 12~\mum.  For 
the reddest three stars in the sample, the S/N stayed above $\sim$100 
out to $\sim$18~\mum.  Each target was observed with the four-point 
dither pattern optimized for point sources in all MRS channels.

\begin{figure}[!ht] % Fig. 3
\includegraphics[width=240pt]{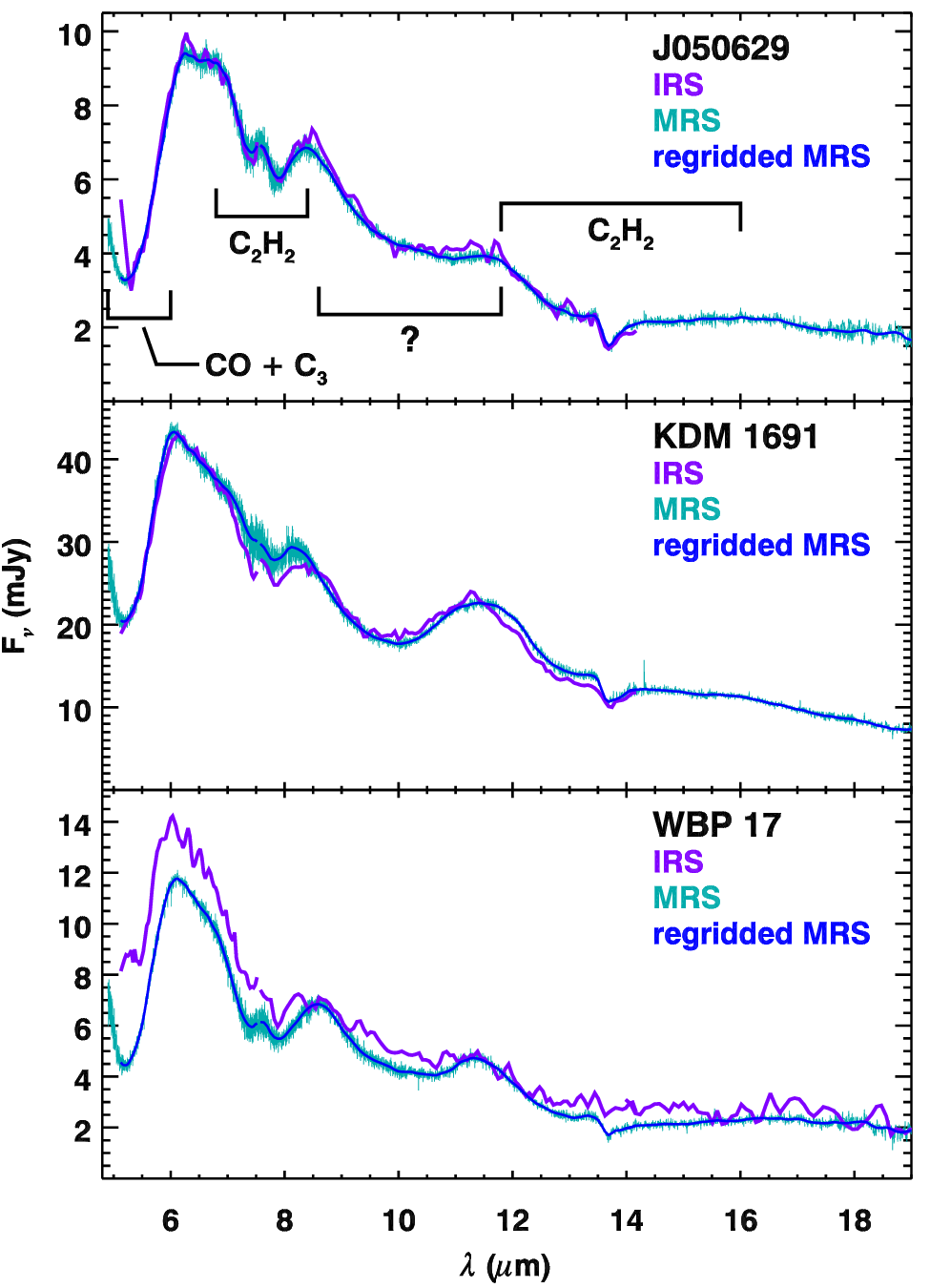}
\caption{MRS spectra of the three SRVs, compared to 
the spectra from the IRS.  The spectra from the MRS are plotted as 
produced by the pipeline and also after regridding them to the 
lower-resolution wavelength grid of the IRS.
\label{f.sp1}}
\end{figure}

\begin{figure}[!ht] % Fig. 4
\includegraphics[width=240pt]{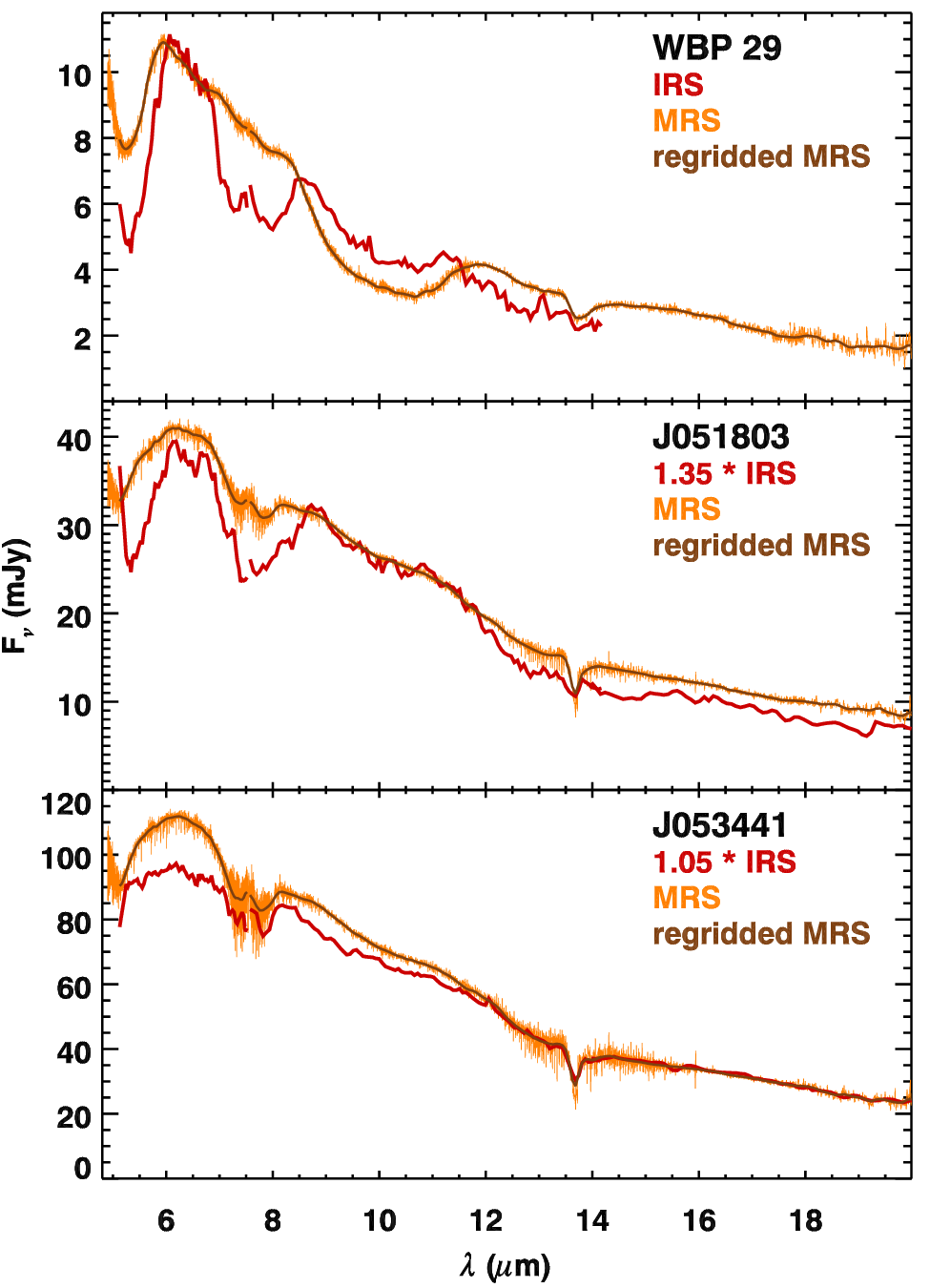}
\caption{MRS spectra of the three bluest of the six Mira variables, 
compared to the spectra from the IRS.  As in Figure~\ref{f.sp1}, the 
spectra from the MRS are plotted before and after regridding to the 
IRS.  Two of the IRS spectra have been scaled up multiplicatively, to 
match the brightness of the MRS spectra, to make comparisons easier.
\label{f.sp2}}
\end{figure}

\begin{figure}[!ht] % Fig. 5
\includegraphics[width=240pt]{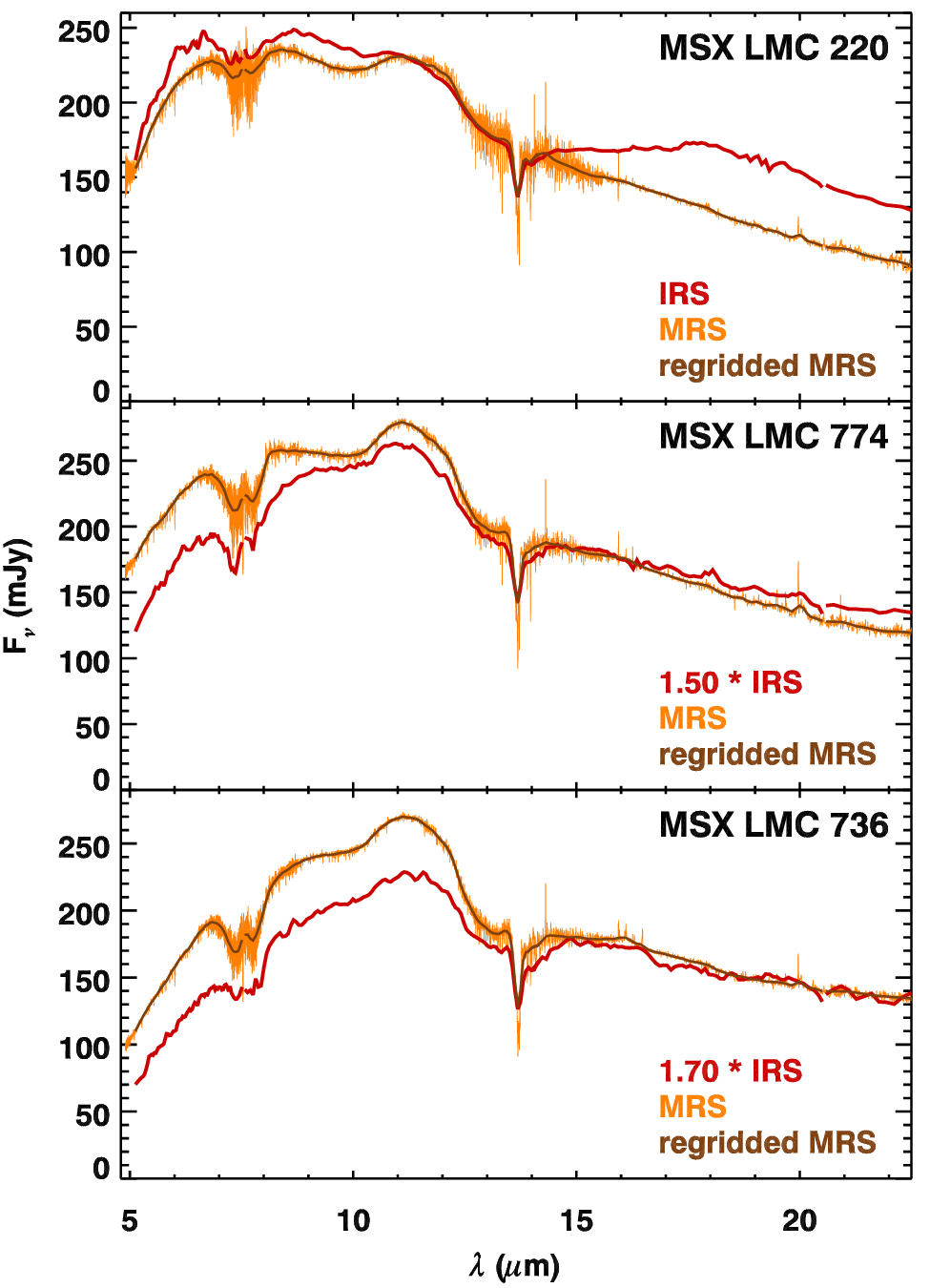}
\caption{MRS spectra of the three reddest Mira variables, compared to 
the spectra from the IRS.  As in Figure~\ref{f.sp2}, the MRS spectra 
are plotted before and after regridding, and two of the IRS spectra 
have been shifted up multiplicatively, to align with the MRS.
\label{f.sp3}}
\end{figure}

The spectral data were processed using the JWST pipeline version 
1.18.0 and Calibration Data Reference System (CRDS) context 
jwst\_1364.pmap.  This version of the pipeline processed the 12 
spectral segments (four channels, three grating segments) separately 
and applied the optional residual defringing algorithm to each.  
The extraction of the spectra used the default pipeline sequence 
that first combines the four dither positions to construct a resampled 
spatial/spectral data cube for each spectral segment.  The pipeline 
then extracts spectra from the data cubes using a circular aperture and 
background annulus with radii that increase linearly with wavelength.
The MRS data cubes showed that our targets dominated their immediate 
environment.  The flux calibration produces results within 1\% of the 
IRS on Spitzer for standard stars observed by both \citep{law24}.  
Figures~\ref{f.sp1} to \ref{f.sp3} present the resulting spectra from 
the MRS and compare them to the IRS.

\section{Spectral Analysis \label{s.analysis}} % Sec. 4

\subsection{The Manchester Method \label{s.mm}} % Sec. 4.1

We have measured the strength of the acetylene absorption band at 
7.5~\mum\ and the SiC dust feature at $\sim$11.5~\mum\ using the 
Manchester Method, which estimates the continuum by fitting a line 
segment to either side of the spectral features.  \citet[][]{slo16} 
describe this method and its history (see their Figures~1--2 for 
illustrations).  While the Manchester Method is arguably simplistic, 
it does provide a quick measurement using a method applied to hundreds 
of other carbon-star spectra for comparison.  It also provides a color 
in two discrete wavelength bands centered at 6.4 and 9.3~\mum, which 
sample spectral regions that come closer to continuum than most others 
in these complex spectra.   The ``continuum'' in these spectra is a 
combination of emission from the star and amorphous carbon dust, which 
has no resonances and thus no spectral features in the infrared.  The 
[6.4]$-$[9.3] color measures the amount of amorphous carbon dust 
around the star, with higher dust-production rates leading to 
increasingly red colors \citep{gro07}.

\begin{deluxetable*}{lrcccc} % Table 2
\tablecolumns{6}
\tablewidth{0pt}
\tablenum{2}
\tablecaption{Spectroscopic Data}
\label{t.spdat}
\tablehead{
  \colhead{Target} & \colhead{[6.4]$-$[9.3]}  & 
  \multicolumn{2}{c}{C$_2$H$_2$ at 7.5~\mum} & 
  \multicolumn{2}{c}{SiC Dust Emission} \\
  \colhead{ } & \colhead{ } &
  \colhead{Wavelength\tablenotemark{a}} & 
  \colhead{Eq.\ width\tablenotemark{b}} &
  \colhead{Wavelength\tablenotemark{a}} & 
  \colhead{SiC/Continuum\tablenotemark{b}\tablenotemark{c}} \\
  \colhead{ } & \colhead{(mag)} & \colhead{(\mum)} & \colhead{(\mum)} & 
  \colhead{(\mum)} & \colhead{ }
}
\startdata
J050629     &    0.144 $\pm$ 0.003 & 7.673 $\pm$ 0.010 & 0.155 $\pm$ 0.001 &
  11.547 $\pm$ 0.011 & (0.097 $\pm$ 0.001) \\
KDM 1691    &    0.031 $\pm$ 0.002 & 7.513 $\pm$ 0.008 & 0.102 $\pm$ 0.001 &
  11.381 $\pm$ 0.006 &  0.238 $\pm$ 0.001 \\
WBP 17      & $-$0.001 $\pm$ 0.004 & 7.460 $\pm$ 0.004 & 0.335 $\pm$ 0.001 &
  11.420 $\pm$ 0.009 &  0.180 $\pm$ 0.001 \\
WBP 29      & $-$0.130 $\pm$ 0.003 & 7.411 $\pm$ 0.023 & 0.012 $\pm$ 0.001 &
  12.165 $\pm$ 0.033 & (0.023 $\pm$ 0.002) \\
J051803     &    0.452 $\pm$ 0.001 & 7.476 $\pm$ 0.005 & 0.106 $\pm$ 0.001 &
  11.079 $\pm$ 0.018 & (0.038 $\pm$ 0.001) \\
J053441     &    0.452 $\pm$ 0.001 & 7.454 $\pm$ 0.003 & 0.122 $\pm$ 0.001 &
  11.266 $\pm$ 0.021 &  0.045 $\pm$ 0.001 \\
MSX LMC 220 &    0.830 $\pm$ 0.001 & 7.538 $\pm$ 0.016 & 0.039 $\pm$ 0.001 &
  11.329 $\pm$ 0.013 &  0.090 $\pm$ 0.001 \\
MSX LMC 774 &    0.908 $\pm$ 0.001 & 7.480 $\pm$ 0.004 & 0.107 $\pm$ 0.001 &
  11.292 $\pm$ 0.008 &  0.141 $\pm$ 0.001 \\
MSX LMC 736 &    1.147 $\pm$ 0.001 & 7.556 $\pm$ 0.008 & 0.133 $\pm$ 0.001 &
  11.325 $\pm$ 0.009 &  0.155 $\pm$ 0.001
\enddata
\tablenotetext{a}{The central wavelength of the band or feature as
defined in the text.}
\tablenotetext{b}{Uncertainties are lower limits; they do not account
for systematic effects, as Section~\ref{s.mm} explains.}
\tablenotetext{c}{Questionable measurements are in parentheses,
as explained in Section~\ref{s.mm}.}
\end{deluxetable*}

\begin{figure}[!ht] % Fig. 6
\includegraphics[width=240pt]{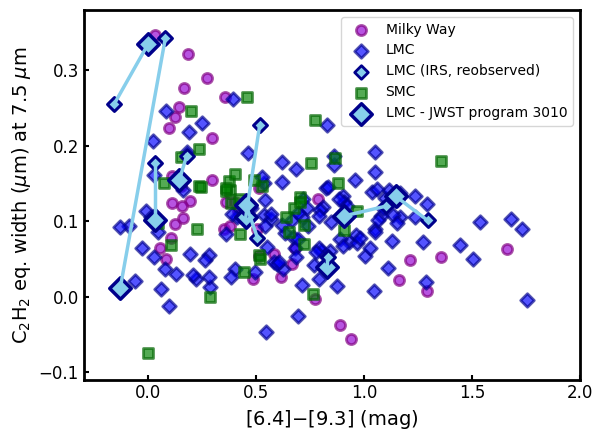}
\caption{The equivalent width of the 7.5~\mum\ acetylene absorption 
band as a function of the [6.4]$-$[9.3] color.  The nine stars in the 
present sample are plotted as small light-blue diamonds, as observed by 
the IRS, and large light-blue diamonds, as observed by the MRS.  WBP~29 
has moved dramatically from top left to bottom left, due to its 
disappearing acetylene band.  Section~\ref{s.c2h2} discusses the
temporal changes at 7.5~\mum\ in WBP~29 and the other sources.
\label{f.c69_w75}}
\end{figure}

\begin{figure}[!ht] % Fig. 7
\includegraphics[width=240pt]{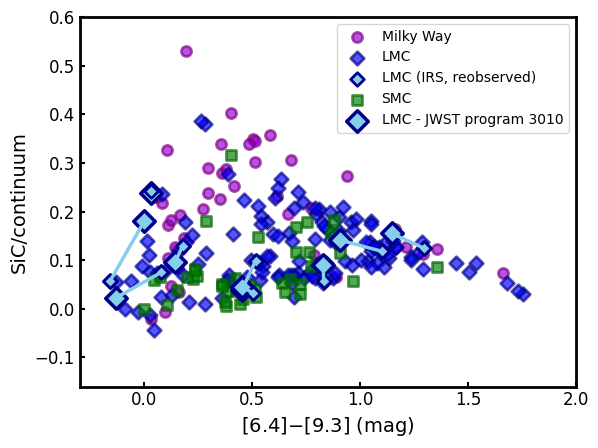}
\caption{The strength of the SiC dust emission feature versus the 
[6.4]$-$[9.3] color, with IRS and MRS results plotted as in
Figure~\ref{f.c69_w75}.
\label{f.c69_sic}}
\end{figure}

Table~\ref{t.spdat} gives the results for the [6.4]$-$[9.3] color, the
equivalent width of the acetylene absorption at 7.5~\mum, and the 
strength of the SiC dust emission feature at $\sim$11.3~\mum\ with 
respect to the continuum.  Figures~\ref{f.c69_w75} and \ref{f.c69_sic} 
plot the strengths of the acetylene absorption and SiC emission versus 
the [6.4]$-$[9.3].  Both figures exclude the eight deeply reddened 
carbon stars observed with the IRS in the LMC with [6.4]$-$[9.3] $>$ 
2.0.

For the absorption band and the dust emission, Table~\ref{t.spdat} 
reports a central wavelength, defined as the wavelength that bisects 
the absorption or emission.  It provides a useful check of the 
extracted feature.  All of the acetylene bands have central 
wavelengths between 7.45 and 7.68~\mum\ and appear to be valid, even 
the weak band in WBP~29.  For SiC dust emission, the central 
wavelength of the extracted feature calls into question the presence 
of SiC in three spectra.  The sample of carbon stars in the LMC 
compiled by \cite{slo16} includes 123 sources with an SiC/continuum 
ratio $>$ 0.05 (out of 184).  For those, the mean central wavelength 
= 11.293 $\pm$ 0.113.  Letting two standard deviations define the 
range of valid central wavelengths gives a range of 11.07--11.52~\mum.  
Both WBP~29 and J050629 are outside this range, leading to doubts that 
their spectra contain SiC dust.  In one other source, J051803, the 
emission feature is close to the blue edge, faint, and therefore also 
questionable.  

The uncertainties in the strengths of the extracted features in 
Table~\ref{t.spdat} are statistical.   They should be treated as lower
limits, because they do not include the systematic effects resulting 
from how spectral structure (from molecular bands or solid-state 
features) on either side of the band or feature considered might 
affect the continuum.  In the case of the SiC dust emission feature, 
absorption from C$_2$H$_2$ to the red will affect the fitted continuum
and thus the measured strength.  As discussed in Section~\ref{s.10um}, 
molecular absorption may also affect the fitted continuum on the blue 
side of the SiC dust.  Thus, a high ratio of feature strength to 
uncertainty does not necessarily mean the feature is real.

\subsection{Analysis of the detailed molecular band structure
\label{s.detailed}} % Sec. 4.2

The analysis uses a grid of synthetic spectra based on
hydrostatic models with effective temperatures of 3100 and 2800~K,
a range of C/O ratios from 1.05 to 4.0, $M$ = 1.0~M$_{\odot}$, 
log~$g$ (cm s$^{-2}$) = 0.0, and a metallicity ([Z/H]) of $-$0.5.  
The models are based on the COMARCS code \citep{ari16, ari19}.  The
models use EXOMOL line lists for C$_2$H$_2$ \citep{chu20}, C$_3$ 
\citep{lyn24}, HCN \citep{har08, bar14}, and CS \citep{pau15}, and 
the HITEMP line list for CO 
\citep{li15}\footnote{\url{https://hitran.org/hitemp}}.  The models 
include isotopologs of the above molecules (at solar abundances), 
other molecules, and atomic transitions.

The hydrostatic models provide synthetic spectra with detailed band
structure for individual molecules, which can greatly aid the
identification of what is absorbing in different wavelength regions.  
The hydrostatic models do not include pulsation, convective 
structures, or winds, and as a consequence they will not include the 
cooler layers of gas likely to be responsible for much of the 
observed molecular spectral structure.  As a result, we should not 
expect the band depths or relative strengths of the available 
transitions in the synthetic spectra to match the observed spectra.  
The present analysis focuses on the detailed spectral structure 
within short wavelength intervals rather than attempting to model the 
overall band strength and shape.  With the exception of measuring 
radial velocities, the analysis is more qualitative than quantitative.

We confine the spectral analysis in this work to what can be 
called the ``flatten-and-fit'' method.  We have broken the 
spectra into multiple wavelength intervals, with intervals at 4.9--5.2 
and 5.2--5.5~\mum\ to investigate the absorption by CO and C$_3$, 
7.0--7.5, 7.5--8.0, and 8.1--8.6~\mum\ for C$_2$H$_2$, HCN, and CS, 
and 12.6--13.1, 13.1--13.6, and 13.8--14.3~\mum\ for C$_2$H$_2$ and 
HCN.  

In each interval, we take the following steps:  (1) create a 
pseudo-continuum from the observed spectrum, by generating a median 
with a 100-pixel interval; (2) divide the spectrum by that 
pseudo-continuum to isolate the spectral structure within the band; 
and (3) subtract 1.0 so that the spectral emission and absorption 
structure oscillates around a mean value of 0.0.  The pseudo-continuum 
runs through the middle of the band structure and not the top, as 
might be more appropriate for an absorption spectrum.  The synthetic 
molecular absorption spectra follow the same steps, except that we 
initially downsampled them to the wavelength grid and resolution of 
the MRS, using the resolutions as defined by \cite{pon24}.  When 
fitting the synthetic absorption to an observed spectrum, we normalize 
the synthetic spectrum to the mean of the absolute value of the 
observed spectrum across a wavelength interval.  The acetylene band at 
7.5~\mum\ flattens pretty well, but the deep and narrow Q branch from 
the $\nu_5$ band at 13.7~\mum\ retains some of its structure, which is 
why the chosen intervals skip from 13.6 to 13.8~\mum.  We placed the 
8~\mum\ interval at 8.1--8.6~\mum, to concentrate on the region where 
CS absorbs most strongly.

\subsubsection{Radial velocities \label{s.vel}} % Sec. 4.2.1

\begin{deluxetable*}{llccccccccc} % Table 3
\tablecolumns{11}
\tablewidth{0pt}
\tablenum{3}
\tablecaption{Radial Velocities}
\label{t.vel}
\tablehead{
  \colhead{Target} & \colhead{$T_{\rm{eff}}$} &
  \multicolumn{8}{c}{Radial Velocity in Wavelength Interval\tablenotemark{a}} &
  \colhead{Mean Radial} \\
  \colhead{ } & \colhead{(K)} & \multicolumn{8}{c} {(km s$^{-1}$)} & 
  \colhead{Velocity\tablenotemark{b}} \\
  \colhead{ } & \colhead{ } & \colhead{4.9--5.2} & \colhead{5.2--5.5} &
  \colhead{7.0--7.5} & \colhead{7.5--8.0} & \colhead{8.1--8.6} &
  \colhead{12.6--13.1} & \colhead{13.1--13.6} & \colhead{13.8--14.3} &
  \colhead{(km s$^{-1}$)} \\
  \colhead{ } & \colhead{ } & \colhead{\mum} & \colhead{\mum} &
  \colhead{\mum} & \colhead{\mum} & \colhead{\mum} & \colhead{\mum} &
  \colhead{\mum} & \colhead{\mum} & \colhead { } 
}
\startdata
J050629     & 3100 & 244 & (184) & 229 & 250 &  240  & (180) & 248 & (263) &
  242 $\pm$  4 \\
KDM 1691    & 3100 & 278 &  270  & 260 & 266 &  261  &  239  & 277 & (183) &
  264 $\pm$  5 \\
WBP 17      & 3100 & 289 &  270  & 261 & 267 &  261  &  238  & 277 &  266  &
  266 $\pm$  5 \\
WBP 29      & 2800 & 240 &  229  & 224 & 240 & (238) & (180) & 228 &  247  &
  235 $\pm$  4 \\
J051803     & 2800 & 241 &  179  & 224 & 204 & (271) & (180) & 223 & (235) &
  214 $\pm$ 11 \\
J053441     & 2800 & 244 & (226) & 224 & 204 & (275) & (180) & 223 & (228) &
  224 $\pm$  8 \\
MSX LMC 220 & 2800 & 240 &  184  & 213 & 203 &  240  & (178) & 206 & (183) &
  214 $\pm$  9 \\
MSX LMC 774 & 2800 & 241 &  221  & 234 & 227 & (276) & (180) & 221 & (183) &
  229 $\pm$  4 \\
MSX LMC 736 & 2800 & 277 &  226  & 224 & 206 & (278) &  189  & 258 & (316) &
  230 $\pm$ 13
\enddata
\tablenotetext{a}{Velocities in parentheses were not used.}
\tablenotetext{b}{Uncertainties are the uncertainties in the mean.}
\end{deluxetable*}

To determine the radial velocities of the stars in the sample, we
used the synthetic spectra from hydrostatic models and the
flatten-and-fit method described in Section~\ref{s.detailed}.   
Table~\ref{t.vel} gives the resulting radial velocities for each 
star in each wavelength interval, as well as the mean.  For the SRVs,
we used the synthetic spectra from models with an effective 
temperature $T_{\rm eff}$ = 3100~K, and for the Miras, with 2800~K.  
For all stars, we used synthetic spectra from models with a C/O 
ratio = 2.0.  Each wavelength interval was fitted separately,
by iterating in 1 km/s steps from 160 to 320 km/s and also through 
different mixtures of the two to three molecules contributing to each 
interval.  Generally, the best radial velocity depended only weakly 
on the details of the chemical mixture.  

This procedure comes with a number of caveats.  First, the 
flatten-and-fit method does not fit the depth of a full absorption 
band; it just fits the fine-scale spectral structure in that band or 
combination of bands.  The only molecules we considered were CO, 
C$_3$, C$_2$H$_2$, HCN, and CS (and not their isotopologs).  Other 
molecules are certainly contributing to the spectra.  As already 
noted, the spectra show evidence of absorption from molecules much 
cooler than one might expect in a hydrostatic model, and that 
difference in temperature could shift the relative line strengths 
within unresolved band structure and thus the apparent radial 
velocity.  We broke broader spectral regions into smaller 
wavelength intervals to reduce the effect of an incorrect 
temperature over larger wavelength ranges.  Finally, the line lists 
may have errors in the predicted line positions and intensities, and 
they may not be complete, which will increase the residuals in the 
fitting process.

All of the limitations just described mean that the systematic
differences between the observed and synthetic spectra dominate the 
noise in the data, which results in a high floor to the $\chi^2$
residuals.  That prevents us from estimating the uncertainty in a
radial velocity fitted to an individual spectral segment.  Instead,
we have estimated the uncertainty in the mean radial velocity for each
star, based on the standard deviations of the radial velocities from 
the separate spectral segments (up to eight).  These velocities can 
vary significantly between intervals for a given star, due in part to 
the caveats already described.  It is also possible that some of the 
differences are real, as different molecules could be tracing 
different layers in the pulsating stellar atmosphere.  Even for the 
same molecule, absorption at different wavelengths could arise from 
gas at different opacities and thus different temperatures and layers.  
Nonetheless, we are able to determine radial velocities with 
uncertainties of less than 10 km/s in most cases.  

Table~\ref{t.vel} presents several radial velocities in parentheses, 
because the fitting method did not produce a satisfactory
minimum. In the 8.1--8.6~\mum\ interval, five of the six Mira variables 
did not show well-defined minima in their $\chi^2$ residuals.  The 
12.6--13.1 and 13.8--14.3~\mum\ intervals were also difficult, leading 
us to use the fitted radial velocities in only the best-behaved cases.  
We rejected two of the fitted velocities at 5.2--5.5~\mum\ 
for the same reason.

Our radial velocities compare well to those from Gaia DR3
\citep{gaia16, gaia23}.  Gaia has radial velocities for three stars 
in our sample.  For KDM~1691, Gaia gives $v_{\rm rad}$ = 260 $\pm$ 
1~km/s, compared to the MRS result of 264 $\pm$ 5~km/s.  For WBP~17, 
$v_{\rm rad}$ = 261 $\pm$ 6 from Gaia and 266 $\pm$ 5 from 
the MRS, and for WBP~29, the corresponding results are $v_{\rm rad}$ = 
240 $\pm$ 3 and 235 $\pm$ 4.  In all three cases, the error bars 
overlap.

\subsubsection{General results \label{s.genresults}} % Sec. 4.2.2

\begin{figure}[!ht] % Fig. 8
\includegraphics[width=240pt]{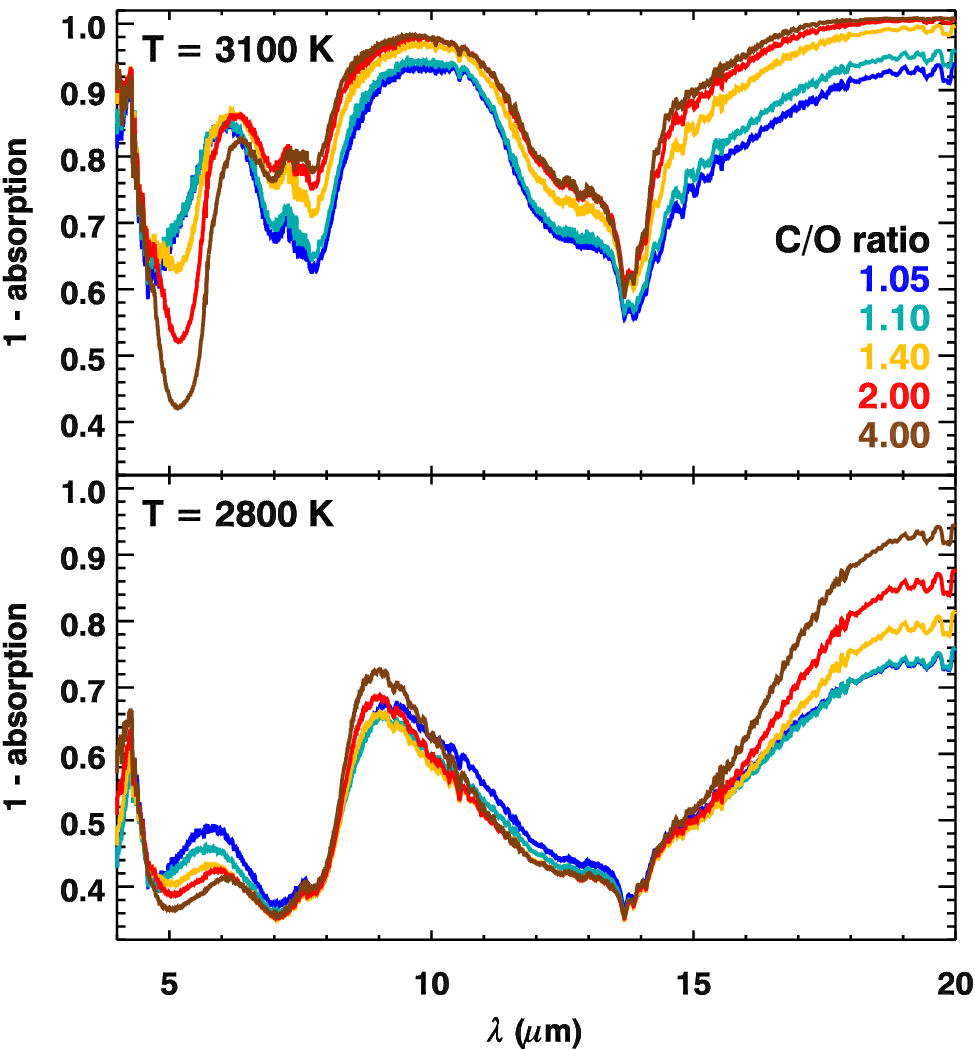}
\caption{The absorption in the synthetic spectra based on hydrostatic
models at two temperatures and five C/O ratios.  The models have been
regridded to the MRS wavelength grid and then smoothed with a 29 pixel 
boxcar.  The 2800 K models show that the spectra are never even close 
to a real continuum in the mid-infrared.
\label{f.model}}
\end{figure}

Figure~\ref{f.model} plots synthetic spectra from the hydrostatic 
models at $T_{\rm eff}$ = 2800 and 3100~K at five different C/O
ratios.  The synthetic spectra are normalized relative to a continuum 
taken from a calculation without any line opacities.  The most 
important takeaway is in the 2800 K models, where the absorption is 
substantial throughout the mid-infrared.  The spectra are nowhere near 
a true continuum level anywhere between 5 and 17~\mum.  That behavior 
illustrates how the Manchester Method is making broad assumptions 
about what can be used for a local ``continuum.''

In most cases in Figure~\ref{f.model}, the acetylene absorption band 
at 7.5~\mum\ in the synthetic spectra does not have the classic ``W'' 
shape seen in the observed spectra.  A comparison of the 
observed and 3100~K synthetic spectra in the top and middle panels of 
Figure~\ref{f.5umlo} shows the difference more clearly.  In the 
hydrostatic models, the C$_2$H$_2$ absorption occurs in a layer with a 
temperature of $\sim$1200--1500 K, but to reproduce the ``W'' shape, 
the absorbing gas needs to be below 1000 K.  \cite{mat06} were able to 
reproduce the band shape with a slab of absorbing C$_2$H$_2$ at 500~K.  
They also noted that the observed C$_2$H$_2$ absorption at 13.7~\mum\ 
in carbon stars heavily obscured by circumstellar dust requires that 
the absorbing gas be above the dust photosphere, which would be in 
the circumstellar envelope and well above the stellar photosphere.  
Hydrostatic models do not reproduce this structure, which is why the 
present analysis is limited to the flattening-and-fitting method.

\begin{deluxetable*}{lllll} % Table 4
\tablecolumns{5}
\tablewidth{0pt}
\tablenum{4}
\tablecaption{Identified Molecules}
\label{t.molecules}
\tablehead{
  \colhead{Target} & \multicolumn{4}{c}{Identified Molecules in 
  Wavelength Range\tablenotemark{a}} \\
  \colhead{ } & \colhead{5~\mum} & \colhead{7--8~\mum} &
  \colhead{8.1--8.6~\mum} & \colhead{13--14~\mum}
}
\startdata
J050629     & \textbf{C$_3$}, CO      & \textbf{C$_2$H$_2$}, HCN, CS &
              \textbf{CS}, C$_2$H$_2$ & \textbf{C$_2$H$_2$}, HCN \\
KDM 1691    & \textbf{C$_3$}, CO      & C$_2$H$_2$, HCN, CS          &
              \textbf{CS}             & \textbf{C$_2$H$_2$}, HCN \\
WBP 17      & \textbf{C$_3$}, CO      & C$_2$H$_2$, HCN, CS          &
              \textbf{CS}, HCN        & \textbf{C$_2$H$_2$}, HCN \\
WBP 29      & \textbf{C$_3$}, CO      & C$_2$H$_2$, HCN, CS          &
              \textbf{HCN}            & \textbf{C$_2$H$_2$}, HCN \\
J051803     & \textbf{C$_3$}, CO      & \textbf{C$_2$H$_2$}, HCN, CS &
              \textbf{CS}             & \textbf{C$_2$H$_2$}, HCN \\
J053441     & \textbf{C$_3$}, CO      & \textbf{C$_2$H$_2$}, HCN, CS &
              \textbf{CS}             & \textbf{C$_2$H$_2$}, HCN \\
MSX LMC 220 & \textbf{C$_3$}, CO      & \textbf{C$_2$H$_2$}, HCN, CS &
              HCN, CS                 & \textbf{C$_2$H$_2$}, HCN \\
MSX LMC 774 & \textbf{C$_3$}, CO      & \textbf{C$_2$H$_2$}, HCN, CS &
              \textbf{CS}, HCN        & \textbf{C$_2$H$_2$}, HCN \\
MSX LMC 736 & \textbf{CO}, C$_3$      & \textbf{C$_2$H$_2$}, HCN, CS &
              \textbf{CS}, HCN        & \textbf{C$_2$H$_2$}, HCN
\enddata
\tablenotetext{a}{Dominant molecules are in \textbf{bold}.  Otherwise,
the order is not significant.}
\end{deluxetable*}

Table~\ref{t.molecules} presents the results of our analysis of the
flattened spectra, where we have fitted the relative molecular 
contributions in the smaller wavelength intervals listed in 
Table~\ref{t.vel} and combined the results into groups of wavelength
intervals.  Where a molecule clearly dominates the absorption, it is
in bold type.  Otherwise, the order of the molecules listed is not 
significant.  Analysis based on more realistic models would be 
required for quantitatively meaningful relative abundances.  
Appendix~\ref{s.flat} plots the best-fitting models in each 
wavelength interval for each star.

\subsubsection{C$_3$ and CO at 5~\mum\ \label{s.5um}} % Sec. 4.2.3

\begin{figure}[!ht] % Fig. 9
\includegraphics[width=240pt]{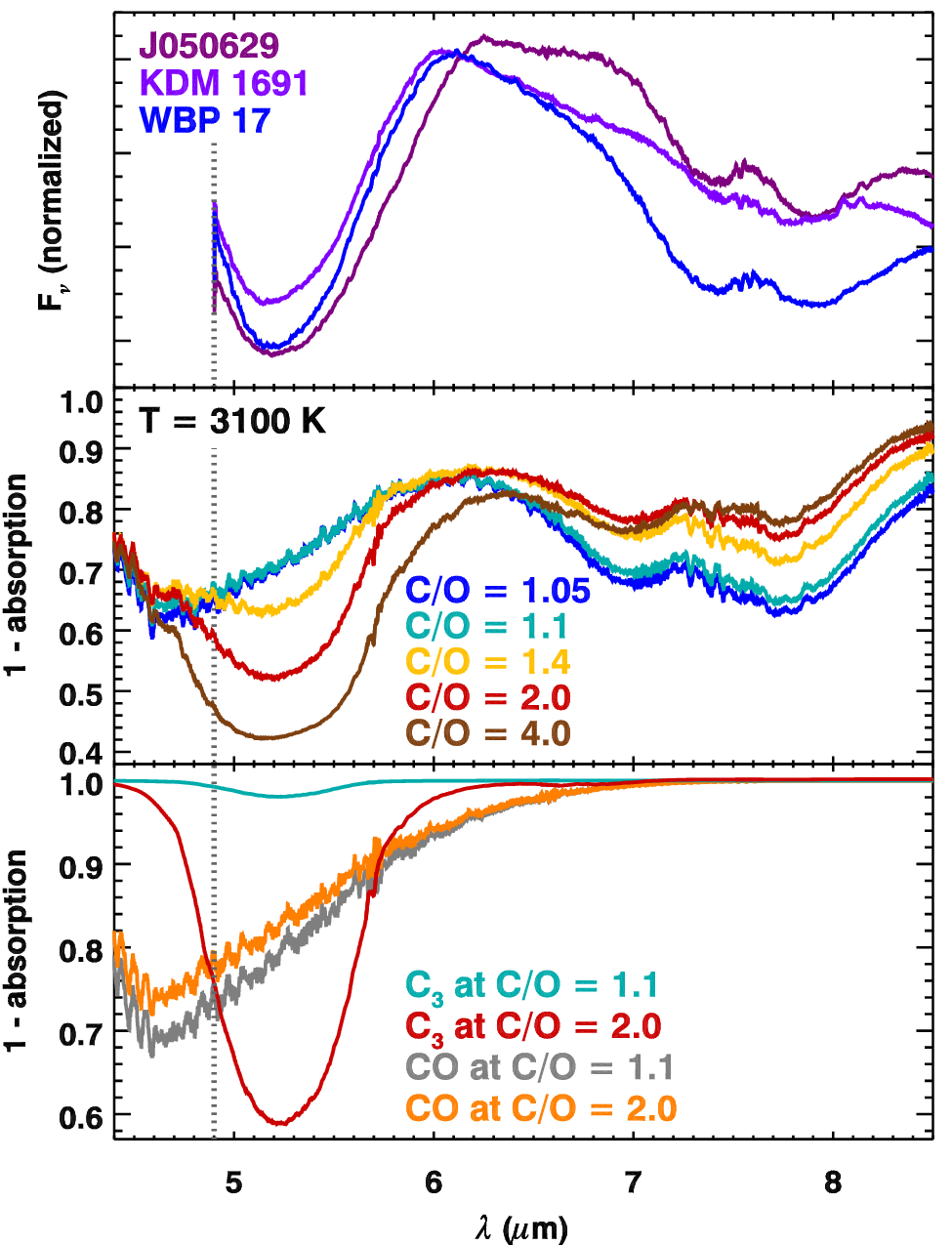}
\caption{Observed spectra of the SRVs compared to absorption in the
synthetic spectra from 4.4 to 8.5~\mum\ at low spectral resolution.  The 
MRS spectra and synthetic spectra have been smoothed with a 29 pixel 
boxcar.  The vertical dotted line at 4.9~\mum\ marks the blue edge of 
the MRS data.  \textit{Top:}  observed MRS data.  \textit{Middle:}  full
absorption of all molecules and atoms considered in the models at 
$T_{\rm eff}$ = 3100~K.  \textit{Bottom:}  absorption from C$_3$ and
CO.  \label{f.5umlo}}
\end{figure}

\begin{figure}[!ht] % Fig. 10
\includegraphics[width=240pt]{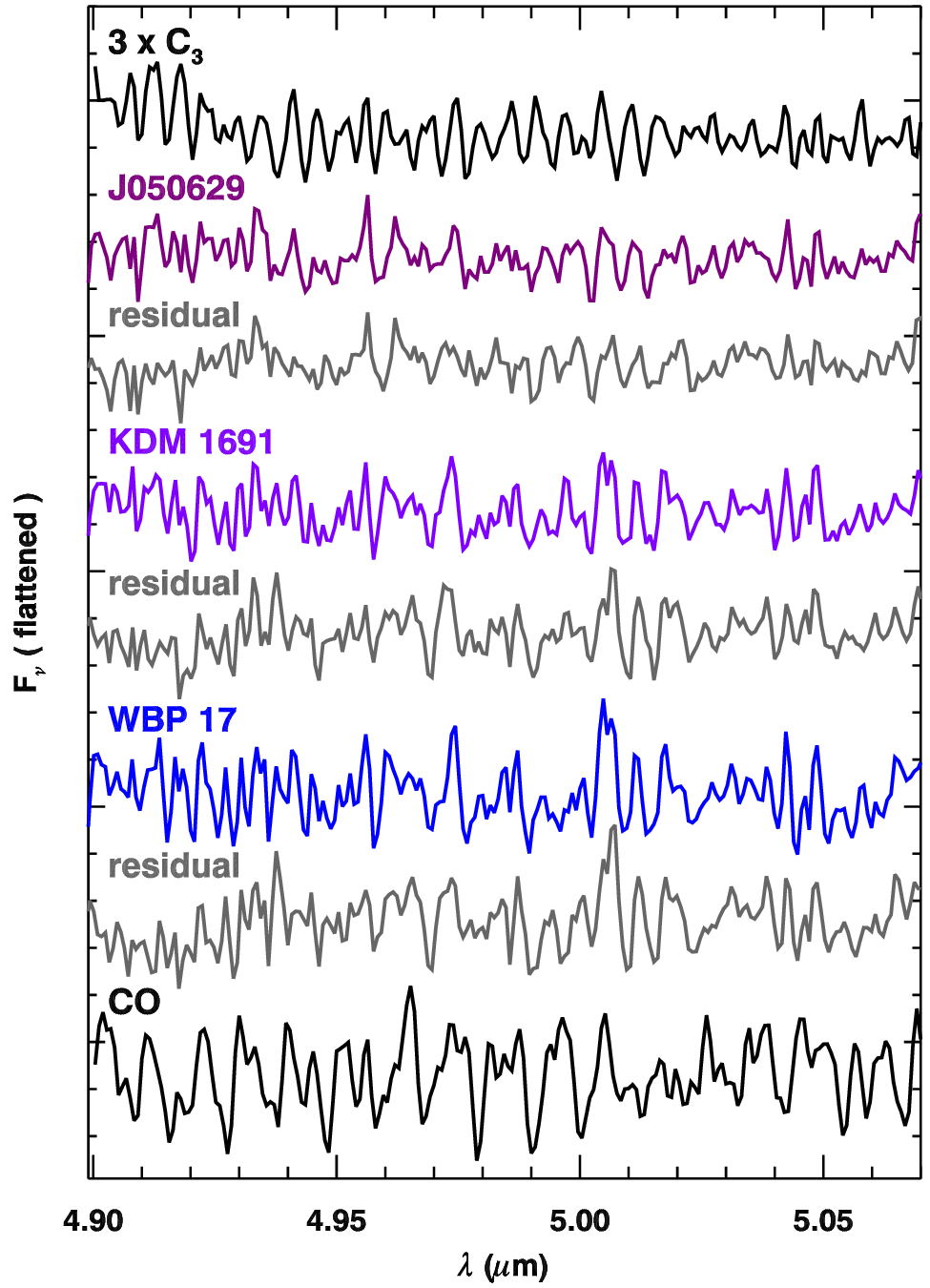}
\caption{A comparison of the band structure of C$_3$ and CO to the
flattened spectra of the SRVs between 4.9 and 5.07~\mum.  The 
synthetic spectra (black, top and bottom) are for 
$T_{\rm eff}$ = 3100 K and C/O = 2.0.  Residual spectra 
after fitting and removing C$_3$ are plotted in gray below the 
spectra for the three sources. \label{f.5umhi}}
\end{figure}

Figure~\ref{f.5umlo} compares the observed spectra of the SRVs with 
synthetic spectra based on 3100 K models after the spectra have been 
smoothed with a 29 pixel boxcar.  The spectra of the SRVs all show a 
minimum at $\sim$5.2~\mum, matching the shape of the C$_3$ 
band in the synthetic spectra.  The CO band, on the other hand, has a 
bandhead further to the blue at $\sim$4.6~\mum.  The synthetic spectra 
show that the C$_3$ band increases in strength as the C/O ratio 
increases, and it also grows wider.  As in Figure~\ref{f.model}, the
absorptions in the synthetic spectra are based on ratios to synthetic
spectra of models computed without line opacities.  

Figure~\ref{f.5umhi} compares the detailed band structure
in the SRVs to absorption from C$_3$ and CO in synthetic spectra based 
on hydrostatic models with $T_{\rm eff}$ = 3100 K and C/O = 2.0.
While the measured $\chi^2$ residuals are smaller for the SRVs 
when using C$_3$ instead of CO, the residual spectra in 
Figure~\ref{f.5umhi} show only a subtle improvement.  This discrepancy 
between the apparent match at low resolution in Figure~\ref{f.5umlo} 
and the higher resolution could arise from the limitations of the 
synthetic spectra used in this analysis.  If the model does not 
reproduce the actual pressure and temperature of the absorbing gas, then
the relative strengths of the individual transitions within the bands 
could differ, which would limit our ability to fit the band structure 
in detail.  Other molecules not considered in our analysis may also
contribute to the spectral structure.  Nonetheless, based on the 
similarity of the shape of the observed and synthetic absorption band 
at 5.2~\mum\ at low resolution, we conclude that C$_3$ is the likely 
carrier of the observed band in the SRVs.  That conclusion results 
primarily from the improved wavelength coverage of the MRS, down to 
4.9~\mum, compared to 5.1~\mum\ with the IRS on Spitzer, which shows 
that the spectra turn up at $\sim$5~\mum, as expected for C$_3$.

The low-resolution spectra from the IRS typically cease to be 
meaningful below 5.1~\mum, while the MRS data are good to 4.9~\mum.
The additional 0.2~\mum\ greatly helps our ability to 
identify C$_3$, as the upturn below the wavelength of peak 
absorption (5.2~\mum) is unambiguous.  
Figures~\ref{f.sp1}--\ref{f.sp3} show that C$_3$ is clearly present in 
the three SRVs and the three blue Miras, but not as obvious in the 
three red Miras.

Comparing the spectra from the MRS to the synthetic spectra 
supports the hypothesis that C$_3$ explains the range of 
reddening in the [5.8]$-$[8] color in the SRVs, which can be seen in 
Figure~\ref{f.sample}.  The synthetic spectra suggest further that the 
cause of the varying quantities of C$_3$ may be the C/O ratio in the 
atmosphere of the SRV.  However, this conclusion is only qualitative,
due to the limits of the hydrostatic models.  The actual C/O values 
needed to reproduce the depth and shape of the observed C$_3$ band 
require further investigation.

\subsubsection{CS at 8~\mum\ \label{s.cs}} % Sec. 4.2.4

\begin{figure}[!ht] % Fig. 11
\includegraphics[width=240pt]{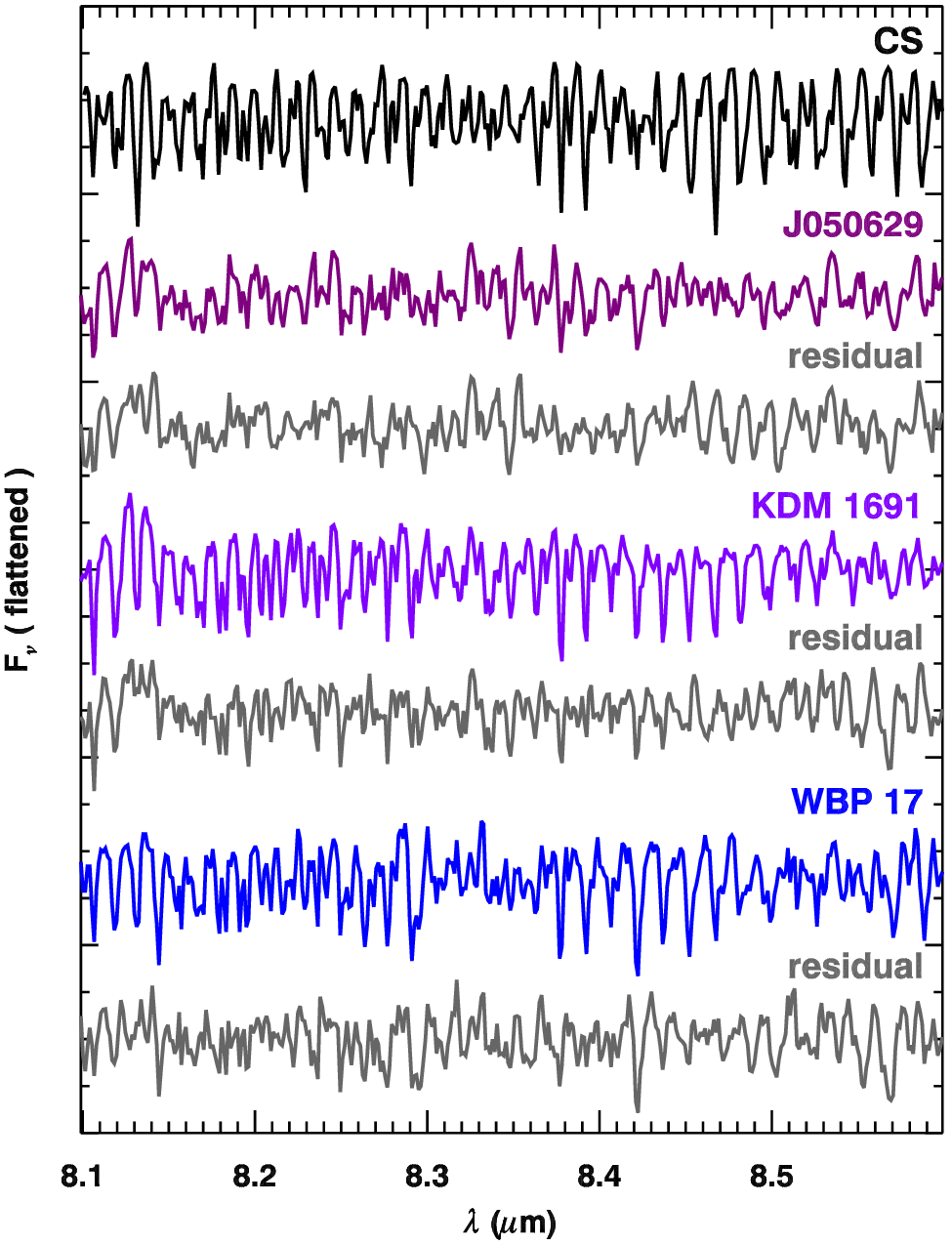}
\caption{The 8.1--8.6~\mum\ spectra of the three SRVs (in color), the 
synthetic spectrum of CS from the hydrostatic model at 3100 K and C/O 
= 2.0 (black), and residual spectra after fitting and removing 
CS (gray).  CS is clearly present in KDM 1691 and WBP 17, and
while less visible in J050629, is required to minimize the residuals.
\label{f.8umcs}}
\end{figure}

Figure~\ref{f.8umcs} compares the 8.1--8.6~\mum\ spectra of the three
SRVs with CS from the synthetic spectrum from a 3100 K model with C/O 
= 2.0.  Including CS reduces the residuals in all three spectra, and
the improvement is particularly clear in KDM~1691 and WBP~17.  The
other possible contributors---C$_2$H$_2$ and HCN---have different
spectral signatures in this wavelength interval, as can be seen in
the spectra in Appendix~\ref{s.flat}.  The spectra in 
Appendix~\ref{s.flat} reveal that CS is present in the SRVs in the 
7.5--8.0~\mum\ interval, too.  Thus, the MRS and its higher spectral 
resolution can detect CS, while the lower resolution of the IRS on 
Spitzer ($R$ $\sim$ 100) could not \citep{mat06}.

\subsubsection{HCN \label{s.hcn}} % Sec. 4.2.5

\cite{mat06} did find evidence of HCN in one or two Magellanic carbon 
stars in their low-resolution IRS spectra.  The spectra from the MRS 
present stronger evidence, but no clear proof like that found for 
CS.  The iterative fitting process of the flattened spectra 
found for all nine stars in the sample that the presence of HCN 
results in lower $\chi^2$ residuals in most or all of the wavelength 
intervals from 7.0 to 14.3~\mum.  The figures in Appendix~\ref{s.flat} 
compare the observed spectra with HCN and other possible contributors.  
Thus, while the evidence for HCN is not conclusive, it is strong.

\section{Discussion \label{s.disc}} % Sec. 5

\subsection{WBP 29 and spectral variations in the sample
  \label{s.c2h2}} % Sec. 5.1

While WBP~29 is classified as a Mira variable, it lies close in
color-color space to the boundary with the SRVs 
(Figure~\ref{f.sample}).  Its pulsational properties separate it 
from the other Miras.  It has the shortest pulsational period of the
Miras, and, for the Miras where we have measured the pulsation 
amplitudes with confidence, it has the lowest amplitude 
(Table~\ref{t.lc} in Appendix~\ref{s.lc}).  

WBP~29 may be in transition from an SRV to a Mira, and its spectrum 
has changed dramatically between the IRS epoch (approaching
minimum) and the MRS epoch (near maximum).  Between those epochs, the 
7.5~\mum\ acetylene band almost disappeared, and an apparent 
10~\mum\ absorption band emerged in its place 
(Figure~\ref{f.sp2}).  \cite{slo24} raised the possibility that the 
changes in its spectral properties could arise from 
evolutionary changes in the star itself, as opposed to changes one 
might expect over the pulsation cycle of the star.  

Since then, we have investigated the infrared light curve 
of WBP~29 more thoroughly (Appendix~\ref{s.lc}), and we have also 
investigated the question of the apparent 10~\mum\ absorption band in 
archival data from the SWS on ISO (Appendix~\ref{s.sws}).  The SWS 
data reveal that the 10~\mum\ band can be seen in the spectra of four 
Galactic carbon stars observed in multiple epochs, and that it 
generally is strongest when the star is at maximum in its pulsation 
cycle, which is also the case for WBP~29.  The relation of 
the 10~\mum\ absorption to stellar pulsation rules out an interstellar 
origin; it must be circumstellar.

With the completion of the observations of the rest of the sample,
we can place WBP~29 in better context.  To examine the behavior of 
the 7.5~\mum\ absorption, we can compare the shift in positions in 
Figure~\ref{f.c69_w75} with the positions of the IRS and MRS epochs 
on the light curves in Appendix~\ref{s.lc}.  We exclude J050629 
from consideration, due to insufficient phase information.  KDM~1691 
was observed by the IRS at minimum and by the MRS at maximum, and its 
7.5~\mum\ band has weakened.  J053441 was observed near minimum by 
the IRS and near maximum by the MRS, and the 7.5~\mum\ band has 
weakened in its spectrum also.  WBP~17 was observed at maximum by 
the IRS and at minimum by the MRS, and, once again, the observation 
at minimum has the stronger absorption at 7.5~\mum.

In J051803, on the other hand, the 7.5~\mum\ band has grown 
slightly stronger in the MRS observation, which was closer to 
maximum, compared to the IRS observation, which was halfway 
from minimum to maximum.  That suggests that the behavior 
of the 7.5~\mum\ depends on more than just pulsation phase.
Similarly, the three reddest Miras present a mixed bag.  All
three were observed at or near minimum by the IRS and by the MRS 
at or near maximum.  However, the changes in the 7.5~\mum\ band
strength are not large, and only two of the three stars
(MSX LMC 220 and 774) follow the trend of stronger absorption at 
minimum.  It is possible in these three redder stars that the 
absorbing layer is much higher above the photosphere and does 
not vary as strongly with the pulsations in the star below it.

To summarize, WBP~29 appears to follow the weak trends of
weaker 7.5~\mum\ absorption and stronger apparent 10~\mum\
absorption close to maximum in its pulsation cycle.  We conclude
that the observed spectral variations most likely result from
its pulsations, as opposed to evolutionary changes.

\subsection{The impact of the apparent 10~\mum\ absorption band 
\label{s.10um}} % Sec. 5.2

\begin{figure}[!ht] % Fig. 12
\includegraphics[width=240pt]{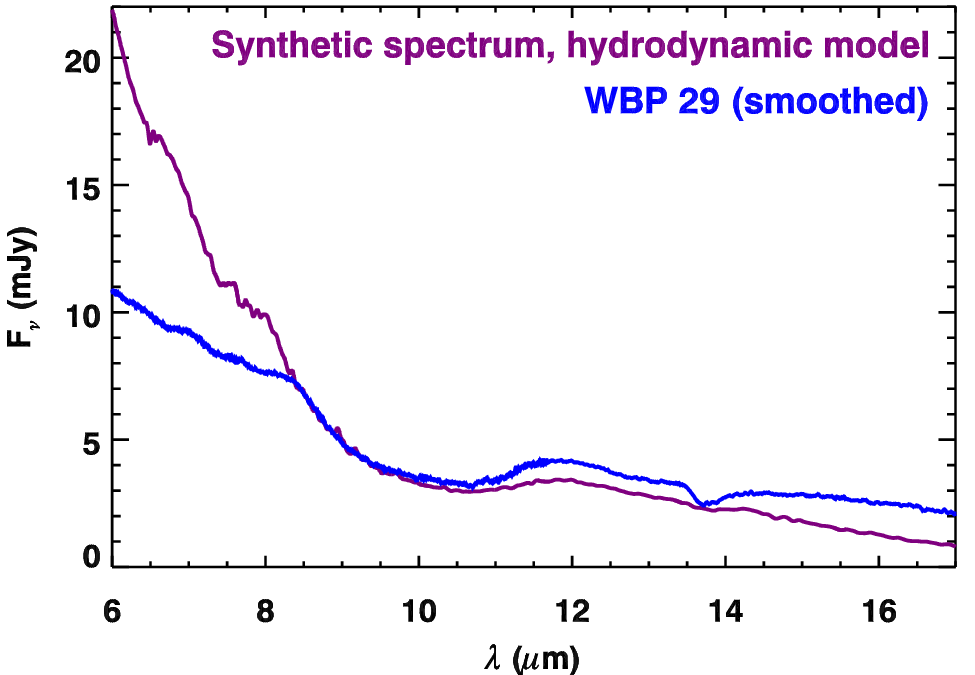}
\caption{The synthetic spectrum of a hydrodynamic model of a carbon 
star, computed without dust opacities and scaled to the spectrum of 
WBP~29.  The synthetic spectrum has been downsampled to the MRS 
wavelength grid and resolution, and the observed spectrum has been 
smoothed with a 15 pixel boxcar.  The model is representative and
not fitted specifically to WBP~29.  It is for a 1~M$_{\odot}$ star
with $T_{\rm eff}$ = 3000 K and $L$ = 7000~L$_{\odot}$.
\label{f.dynmodel}}
\end{figure}

The spectra of the SRVs and the blue Mira WBP~29 in Figures~\ref{f.sp1}
and \ref{f.sp2} show a clear depression centered at 10~\mum, which 
could be a molecular absorption band.  The depth of this feature and 
the position of its long-wavelength edge may vary between objects and 
with time.  As Appendix~\ref{s.sws} shows, some Galactic carbon stars 
observed by the SWS behave similarly.

The carrier of the apparent absorption band at 10~\mum\ is uncertain.  
We were unable to match the spectral structure in that region closely 
to any of the molecules considered in the synthetic spectra based on 
hydrostatic models (viz C$_2$H$_2$, HCN, CS, CO, and C$_3$).  An 
unidentified molecule not included in our models could be responsible.

Another possibility is that the apparent absorption at 10~\mum\ arises 
from molecular emission to either side, at 7.5 and 12--15~\mum,
most likely from C$_2$H$_2$, with possible contributions from HCN.  
These molecules have millions of overlapping transitions in this 
spectral region, which could produce a raised quasi-continuum in the 
right conditions.  Those conditions could be met if the pulsations in 
the carbon stars are strong enough to push a layer of gas 
high above the stellar photosphere.

Figure~\ref{f.dynmodel} plots a synthetic spectrum from a snapshot
of a hydrodynamic model with an extended pulsating atmosphere and a
dust-driven wind, alongside a spectrum of WBP~29.  This model is
based on the DARWIN code \citep{hof16} and is similar to the
carbon-star models by \cite{eri23} and \cite{sid25}, except for the
updated molecular opacities, as in the hydrostatic COMARCS models
(see Section~\ref{s.detailed}).  While the dynamic simulations
include dust, we have computed the synthetic spectrum without dust
opacities.  That choice makes the synthetic spectrum bluer, but it 
also removes dust as a possible cause of the apparent absorption at 
10~\mum.  In the synthetic spectrum, the apparent dip at 10~\mum\ is 
just continuum between molecular emission from other molecules to 
the red and to the blue.  The 7.5~\mum\ band can be seen in emission 
in Figure~\ref{f.dynmodel}, as can the broad $\nu_5$ C$_2$H$_2$ band 
with its Q branch at 13.7~\mum\ and the P and R branches extending 
the band to $\sim$12 and 15~\mum.  

The similarity of the shape of the synthetic and observed spectra in
the 10~\mum\ region demonstrates that the emission scenario is
plausible, but direct detailed evidence for emission on either side
of the 10~\mum\ feature is lacking.  In WBP~29, C$_2$H$_2$, HCN, and
CS are all detected between 7.0--7.5~\mum, but in absorption.  The
absorption could arise from a cooler layer in the line of sight above
the emitting layer.  The same argument could also apply at 13.7~\mum, 
where the $\nu_5$ Q branch of C$_2$H$_2$ is clearly in absorption.

More work is needed on the nature of this apparent
absorption band at 10~\mum.  If it is an absorption band,
its presence is affecting the [6.4]$-$[9.3] color, which is used as a 
proxy for the ratio of dust to stellar emission and the 
dust-production rate.  In Figure~\ref{f.c69_w75}, the strong dip in 
the spectrum of WBP~29 at 10~\mum\ appears to have driven the 
[6.4]$-$[9.3] color negative.  Future work could mitigate this by 
shifting the 9.3~\mum\ band to a shorter wavelength, perhaps 
$\sim$8.7~\mum, which would move it partly out of the 10~\mum\ dip.

Measurements of the SiC dust emission feature are also affected
by the 10~\mum\ dip to the blue, as well as absorption from C$_2$H$_2$
and other molecules to the red.  At first glance, the synthetic 
spectrum in Figure~\ref{f.dynmodel} might look like it has SiC dust
emission at $\sim$11.3~\mum, but it does not.  \cite{lei08} 
tried to address this challenge by modeling the 
``continuum'' (from the star, amorphous carbon dust, and molecular 
absorption) when measuring the flux from SiC dust.  A simpler approach 
would be to focus on the centroid of the extracted feature, as done in 
Section~\ref{s.mm}.  While that might not remove all false positives, 
it would certainly catch some.

\subsection{MSX LMC 220 and Its Vanishing Dust Feature} % Sec. 5.3

The dust emission in the spectrum of MSX~LMC~220 has changed 
dramatically between the IRS and MRS epochs (Figure~\ref{f.sp3}).  
The IRS spectrum shows a cool dust emission feature that peaks in the 
vicinity of 18~\mum.  We examined the spectral images from which the 
IRS spectrum was extracted and saw no evidence for a contaminating 
source or anything else unusual, leaving us to conclude that the 
earlier spectrum really does show this extra dust component.  This 
feature could be related to an 18~\mum\ shoulder seen in the dust 
emission from some carbon-rich planetary nebulae by \cite{ber09}, who 
noted that while it was in roughly the right position to be the 
O--Si--O bending mode in silicate dust, one would expect to see the 
Si--O stretching mode at 10~\mum, which is absent, and oxygen-rich 
dust would be unlikely in a carbon-rich environment.  They were unable 
to identify the carrier of the feature.

The disappearance of this apparent dust emission in the time between
the IRS and MRS epochs is as much of a mystery as its origin in the
first place.  A simplistic calculation assuming blackbodies for the
grains, radiative equilibrium in a stellar radiation field
of $T_{\rm eff}$ = 3000 K for MSX~LMC~220, and an outflow velocity of 
10 km/s, results in changes in dust temperature too small for it
to vanish from the infrared spectrum in the intervening 19.6 yr.  
For example, a grain at 150 K when observed by the IRS would have 
cooled to 140 K by the MRS epoch.  A grain temperature as high as 
1000~K is highly unlikely, but even then it would have cooled only to 
380~K.  MSX~LMC~220 is the most luminous source in our sample, which 
raises the possibility that it is an unusual object and not a normal 
star on the AGB.

\subsection{Carbon Stars in the Early Universe\label{s.context}} % Sec. 5.4

The MRS enables the detailed study of the molecular gas in the 
envelopes of carbon stars, and that probes the chemistry of the gas 
from which carbon-rich dust condenses.  
The introduction noted that carbon stars likely contribute newly
created dust to the early Universe, but they are often disregarded. 
Claims that the first carbon stars could not form until $\sim$1~Gy 
after the Big Bang (i.e., at a redshift of $\sim$5.7) have circulated
for years \citep[e.g.,][]{tod01, dwe07}.  Instead, the alternative 
explanation has been supernovae \citep[e.g.,][]{mai04, dwe07}.  
However, the 1~Gyr delay for carbon star formation is outdated.
Typically, $\sim$500 Myr are allowed for the first stars to form and 
another 500 Myr for carbon stars to form on the AGB.  Both of those 
timescales are overestimates.

Regarding the first timescale, JWST has detected galaxies at $z$ $>$ 
14, and spectra reveal emission from the optical [O III] doublet 
shifted into the mid-infrared \citep[e.g.,][]{hel26, nai26}.  Since 
oxygen requires stellar nucleosynthesis, stars must have formed 
within 300 Myr of the Big Bang.

Regarding the second, stellar models show that metal-poor stars reach 
the AGB more quickly than their metal-rich analogs.  For example, 
\cite{vw93} found that at Solar metallicity, or $Z$ $\sim$ 0.016, a 
2~M$_{\odot}$ star takes $\sim$1.6~Gy to evolve from the zero-age main 
sequence to the thermal-pulsing AGB, while the same star with $Z$ 
$\sim$ 0.004 takes 1.2~Gyr.  \cite{her04} found that at $Z$ = 0.0001, a 
2~M$_{\odot}$ star takes 800~Myr.  At 5~M$_{\odot}$, the corresponding 
times are 123, 102, and 97~Myr.

A key question, though, is what the upper mass limit is for carbon
stars at low metallicities.  The cutoff is driven partly by proton 
capture at the bottom of the convective envelope, known as hot bottom 
burning, which converts carbon to nitrogen \citep{ren81}.  
\cite{ibe83} noted that the creation of a carbon star on the AGB 
depends on the interplay between helium burning, dredge-up, proton 
capture, and details of the convection.  More recently, the models of
\cite{str23} suggest that the upper mass limit increases from
$\sim$3~M$_{\odot}$ at Solar metallicity to $>$ 5~M$_{\odot}$ at 
$Z$ = 0.002.  Although determining the upper mass limit for carbon 
stars remains a challenge, this work suggests that carbon stars could 
form early in the life of the Universe.

Combining our knowledge of how soon stars can form, 300~Myr, with the 
time a metal-poor 5~M$_{\odot}$ star takes to reach the 
thermal-pulsing AGB, 100 Myr, carbon stars could reasonably be expected 
to appear, and start creating dust, as early as 400~Myr after the Big 
Bang, or at a redshift of 11.4.  Based on observations of metal-poor 
carbon stars in the Local Group \citep{slo12, slo16, boy25} those 
earliest metal-poor carbon stars likely produced significant amounts 
of carbon-rich dust and ejected it back into their host galaxies.

\subsection{Molecular chemistry and metallicity\label{s.chem}} % Sec. 5.5

This paper represents an initial assessment of the quality of the 
spectra and what they can reveal.  The spectra are a complex blend of 
overlapping absorption bands from a multitude of carbon-bearing 
molecules, probably absorbing from multiple layers with different 
physical conditions, making them a challenge to model.  The 
rewards of that modeling, though, promise to be rich.  We can use
chemical models and abundances of molecules like HCN and C$_3$ to
determine the C/O ratio and probe isotopic ratios 
(e.g., $^{13}$C/$^{12}$C and $^{15}$N/$^{14}$N) in gas dredged up 
from the nuclear-burning interior.

This paper has focused on the first results from a simplified spectral 
analysis method described as ``flatten-and-fit.''  While this method
avoids the complexities of detailed modeling, it comes with a 
number of caveats (as discussed in Section~\ref{s.detailed}) which 
limit how quantitative conclusions can be.  Nonetheless, it has 
strengthened the case for absorption from C$_3$, CS, and HCN.  The 
presence of C$_3$ may require a shift in thinking from C$_2$H$_2$ as 
the only molecule dominating the mid-infrared absorption.  The 
presence of CS and HCN point to the complexity of stellar enrichment 
at the metallicity of the LMC and raise the question of whether the
MRS can detect these molecules at even lower metallicities.

Future work on this dataset will include more sophisticated analyses,
to better disentangle the band structure and model the gas.  That will 
probe the temperature and pressure of the absorbing molecules and 
better constrain the models.  Improved understanding of the chemical 
and physical conditions in the circumstellar envelopes of the carbon 
stars will lead to improved understanding of the initial conditions 
for dust condensation.  The current results for these nine carbon 
stars in the LMC demonstrate the capabilities of JWST's MRS to probe 
their properties more deeply than previous instruments and to reach 
even more metal-poor carbon stars that more closely resemble the 
carbon stars forming at high redshift.

\section{Summary\label{s.sum}} % Sec. 6

Observations of nine carbon stars in the LMC with the MRS 
reveal detailed molecular structure in their spectra.  
C$_2$H$_2$ produces strong and broad absorption bands centered at 7.5 
and 13.7~\mum\ that were well known before.  
Comparing the observed spectra structure to synthetic spectra 
from hydrostatic models of carbon stars reveals that CS is
absorbing at $\sim$8~\mum.  The evidence for HCN is indirect, but 
including it in the fitted mixtures of absorptions from synthetic 
spectra in the wavelength intervals with C$_2$H$_2$ generally results 
in lower residuals in the fitting process.  The MRS and 
synthetic spectra also support the identification of C$_3$ as the 
carrier of an absorption band centered at 5.2~\mum, a result made 
possible by the improved wavelength coverage below 5~\mum\ in the 
MRS compared to the IRS on Spitzer.  The presence of C$_3$ explains 
the range of [5.8]$-$[8.0] colors observed in SRVs.

WBP~29 shows spectral structure that looks like a strong and broad 
10~\mum\ absorption band in the MRS data that is not apparent in the
IRS epoch.  This spectral structure appears in other spectra 
in our sample.  A review of multi-epoch spectroscopy from the SWS on 
ISO confirms its presence and its variability in Galactic carbon 
stars.  If the apparent 10~\mum\ dip is an absorption band, its 
carrier is unknown.  Hydrodynamic models of carbon stars offer an 
alternative explanation, that we may be observing a continuum at 
10~\mum\ bracketed by emission at shorter and longer wavelengths.  The
variation appears to be related to the pulsation cycle of the star,
with a stronger ``dip'' at 10~\mum\ when the star is at optical
maximum.

This variable spectral structure in the vicinity of the SiC dust
emission feature at $\sim$11.3~\mum\ raises some concern about the
reliability of the measured dust emission strengths using the Manchester
Method.  That approach fits line segments above absorption bands 
and below emission features to estimate the continuum, but these
spectra have no real continuum in the classic sense.  For strong
SiC emission, these problems are less significant, but detections
of fainter SiC dust features may not be real.  They may simply
be continuum between absorption bands or shoulders between emission
and absorption bands.  We recommend that users of the Manchester
Method use the measured central wavelengths of bands and features
to identify possible false positives.  It may also make sense 
to shift some of the continuum positions, such as the wavelengths used 
to measure the [6.4]$-$[9.3] color, which has served as a proxy for 
the amount of amorphous carbon dust contributing to the spectrum.

The spectrum of MSX~LMC~220 changed significantly between the
IRS and MRS epochs, with the older data showing a broad emission
feature from apparent dust at $\sim$18~\mum\ that is not present
in the MRS data.  The origin of the feature in the previous epoch
and the reason for its disappearance are not known.

\newpage

\begin{acknowledgments}

The authors thank the anonymous referee for thoughtful feedback
that resulted in an improved manuscript.
Support for G.C.S., K.E.K., E.J.M., and R.S.\ was provided through 
grant JWST-GO-03010.005 under NASA contract NAS5-03127.
B.A.\ and S.H.\ acknowledge funding from the European Research Council 
(ERC) under the European Union's Horizon 2020 research and innovation 
program (grant agreement No.\ 883867, project EXWINGS) and the 
Swedish Research Council (Vetenskapsradet, grant no.\ 2019-04059).
J.C.\ acknowledges support from the University of Western Ontario and the 
Natural Sciences and Engineering Research Council of Canada.
K.J.\ is supported by the Swedish National Space Agency, and M.M.\ is
supported by STFC consolidated grant (ST/W000830/1).  
I.M.\ and A.A.Z.\ acknowledge funding from the OSCARS project, which has 
received funding from the European Commission's Horizon Europe Research 
and Innovation program under grant agreement No.\ 101129751.
The contribution of R.S.\ to the research described here was carried 
out at the Jet Propulsion Laboratory, California Institute of 
Technology, under a contract with NASA (80NM0018D0004).

We gratefully acknowledge the observations of variable stars from
the International Database of the American Association of Variable
Star Observers (AAVSO) contributed by observers around the world.  
Some of the observations used here were obtained by the British 
Astronomical Association for Variable Star Section.
This research has made use of the VizieR catalog access tool at 
the Strasbourg Astronomical Data center (CDS) in Strasbourg, France 
\citep{och00}, the Astrophysics Data System (ADS) funded by NASA under 
Cooperative Agreement 80NSSC25M7105 at the Smithsonian Astrophysical 
Observatory, and the Infrared Science Archive (IRSA) operated by IPAC
at the California Institute of Technology.  It has also utilized data 
from the Gaia mission of the European Space Agency (ESA;
\url{https://www.cosmos.esa.int/gaia}), processed by the Gaia Data
Processing and Analysis Consortium (DPAC) and funded by the
institutions participating in the Gaia Multilateral Agreement.

Coauthor Paola Marigo died while this program was still obtaining 
observations.  Her work strongly contributed to the success of our 
observing proposal, and we deeply regret that we did not have the
opportunity of analyzing and interpreting the spectra alongside her.

\end{acknowledgments}

\vspace{5mm}

\facilities{AAVSO, ISO (SWS), Spitzer (IRAC, IRS), WISE, JWST (MIRI, MRS)}

\appendix

\section{Infrared light curves \label{s.lc}} % Appendix A

We can use multiepoch infrared photometry to determine the phase of
the pulsational cycle of the stars when the IRS and MRS 
obtained spectra.  We constructed light curves for all nine targets 
using 3.6 and 4.5~\mum\ photometry from the SAGE and SAGE-Var surveys 
and 3.4 and 4.6~\mum\ photometry from WISE.  Most of the targets have 
29 epochs, two from SAGE in 2005, four from SAGE-Var in 2010 and 2011, 
two from the original WISE mission in 2010, and 21 from the reactivated 
WISE mission, NEOWISE-R, which ran from 2014 to mid-2024.

The WISE epochs are spaced roughly 6 months apart.  Each of these
epochs consists of many individual scans obtained every few hours.  
Because the LMC is close to the south ecliptic pole, targets in the
sample had an average of between 174 and 391 of these individual
scans per epoch.  The photometry from the individual scans was 
combined by finding a median, filtering out any data more than 0.35 
magnitudes from that value, then redetermining the median.

The SAGE and SAGE-Var data provide epochs that are not on the 
6 month cadence of the WISE data, so they can help break 
degeneracies.  The SAGE-Var data were taken contemporaneously with the 
first two original WISE epochs, making them particularly helpful, 
and the SAGE epochs in 2005 extend the photometric baseline to 19 yr.

\begin{deluxetable*}{lrr} % Table 5
\tablecolumns{3}
\tablewidth{0pt}
\tablenum{5}
\tablecaption{Color and Magnitude Corrections for Carbon Stars}
\label{t.corr}
\tablehead{ \colhead{Relation} & \colhead{y-intercept} & \colhead{Slope} }
\startdata
  (W1$-$W2) $-$ ([3.6]$-$[4.5]) versus [3.6]$-$[4.5] &    0.1528 & 0.3003 \\
  W1$-$[3.6] versus [3.6]$-$[4.5]                    &    0.0352 & 0.3449 \\
  W2$-$[4.5] versus [3.6]$-$[4.5]                    & $-$0.0224 & 0.0537
\enddata
\tablecomments{These corrections are only valid for [3.6]$-$[4.5] $<$ 1.55.}
\end{deluxetable*}

The IRAC data require a correction to align with the WISE data.  The first 
part of the correction is based on the color corrections generated by 
\cite{slo16}, except that they converted from WISE to IRAC.  We repeated 
their analysis, going in the other direction and adding the WISE epochs 
since 2016 to the analysis.  Table~\ref{t.corr} gives the results, based 
on least-squares fits of lines to the mean magnitudes for all of the 
carbon stars in the Spitzer/IRS sample defined by \cite{slo16}.  That 
sample includes 184 carbon stars in the LMC and 40 in the SMC.  Stars 
with [3.6]$-$[4.5] colors $>$ 1.55 were excluded, because the color 
corrections for the reddest carbon stars do not appear to follow any 
coherent trend.

The corrections in Table~\ref{t.corr} leave the IRAC data still offset
from the WISE data by typically $\sim$0.15--0.25 mag.  These offsets 
are within the scatter of the photometry for the full sample \cite[see 
Figure~17 from][]{slo16}, and they are generally unimportant when 
fitting the large-amplitude variables at the red end of the Mira 
sequence.  For weaker pulsations, though, the offsets make using the 
IRAC data challenging.  

To remove these residual offsets, we first fitted a sinusoid to just 
the WISE photometry, then used that to determine the mean residuals 
for the IRAC photometry after the first correction was applied.  The 
second correction to the IRAC photometry then removed those offsets 
source by source.  As explained below for individual stars, two stars 
show long-term variations in their light curves in addition to the 
fitted sinusoid, and for those we forced the average residuals for the 
four SAGE-Var and two original WISE epochs, all taken in 2010--2011, 
to match.

Our algorithm for fitting light curves is based on the one used by 
\cite{slo16}.  We iterated through pulsation periods, typically from
100 to 1200 days, with a step size of 1 day, fitted a sine function 
to the W1 and W2 data independently, and picked the period with the 
least $\chi^2$ residuals for each.  The pulsation periods for W1 and 
W2 were then averaged, and sinusoids were fitted to determine the 
mean magnitudes and amplitudes for W1 and W2.  That result served as 
the basis for determining the second correction to the IRAC 
photometry.

With the IRAC photometry corrected to be consistent with WISE, we then 
fitted sinusoids to the combined IRAC and WISE photometry, again 
iterating from 100 to 1200 days and searching for the period with the 
minimum residuals.  Table~\ref{t.lc} gives the results.  The reported 
period is the average from fitting W1 and W2 separately, and the mean 
magnitudes and amplitudes were determined by using that mean period.  
The amplitudes are peak-to-peak for the fitted sinusoid.  The 
uncertainty in period is the standard deviation of the four periods
found for W1 and W2 with and without using the IRAC photometry,
illustrating how the measured periods depend on the data available.
The zero-phase epoch is the average of the results for W1 and W2 in
the final fitting step (fixed period, IRAC and WISE data), and the
quoted uncertainty is the uncertainty in the mean (i.e., half the
difference).  

\begin{figure}[!ht] % Fig. 13
\includegraphics[width=240pt]{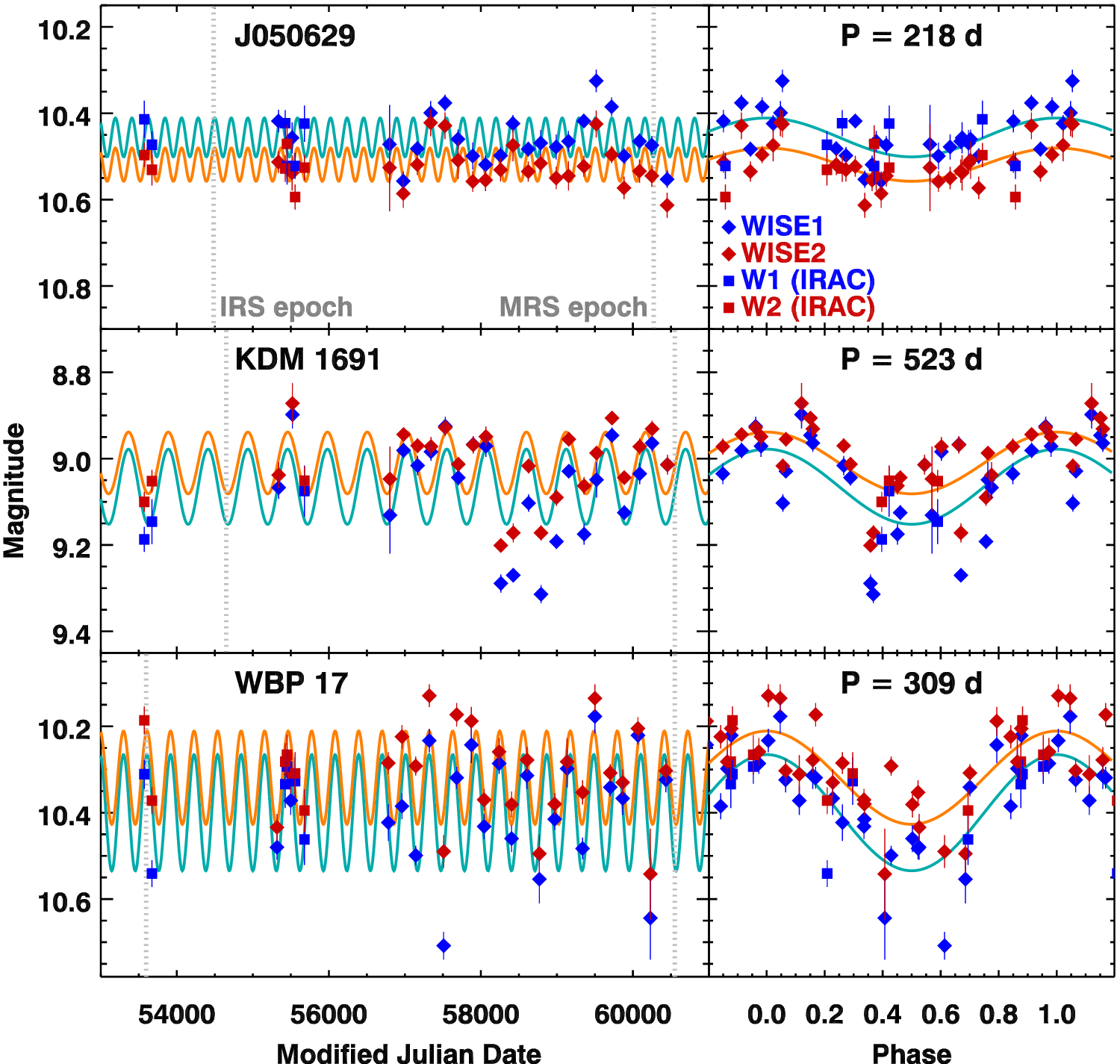}
\caption{Light curves for the three SRVs in the sample, plotted versus
MJD (left) and phased (right).
IRAC data color-corrected to the WISE filters are plotted as squares.
The fitted sine functions are plotted in light blue for W1 and orange 
for W2.  In the left-hand panels, the vertical dashed lines give the 
epochs of the IRS and MRS observations.  All three panels have the same 
vertical range, 0.75 mag.
\label{f.lc1}}
\end{figure}

\begin{figure}[!ht] % Fig. 14
\includegraphics[width=240pt]{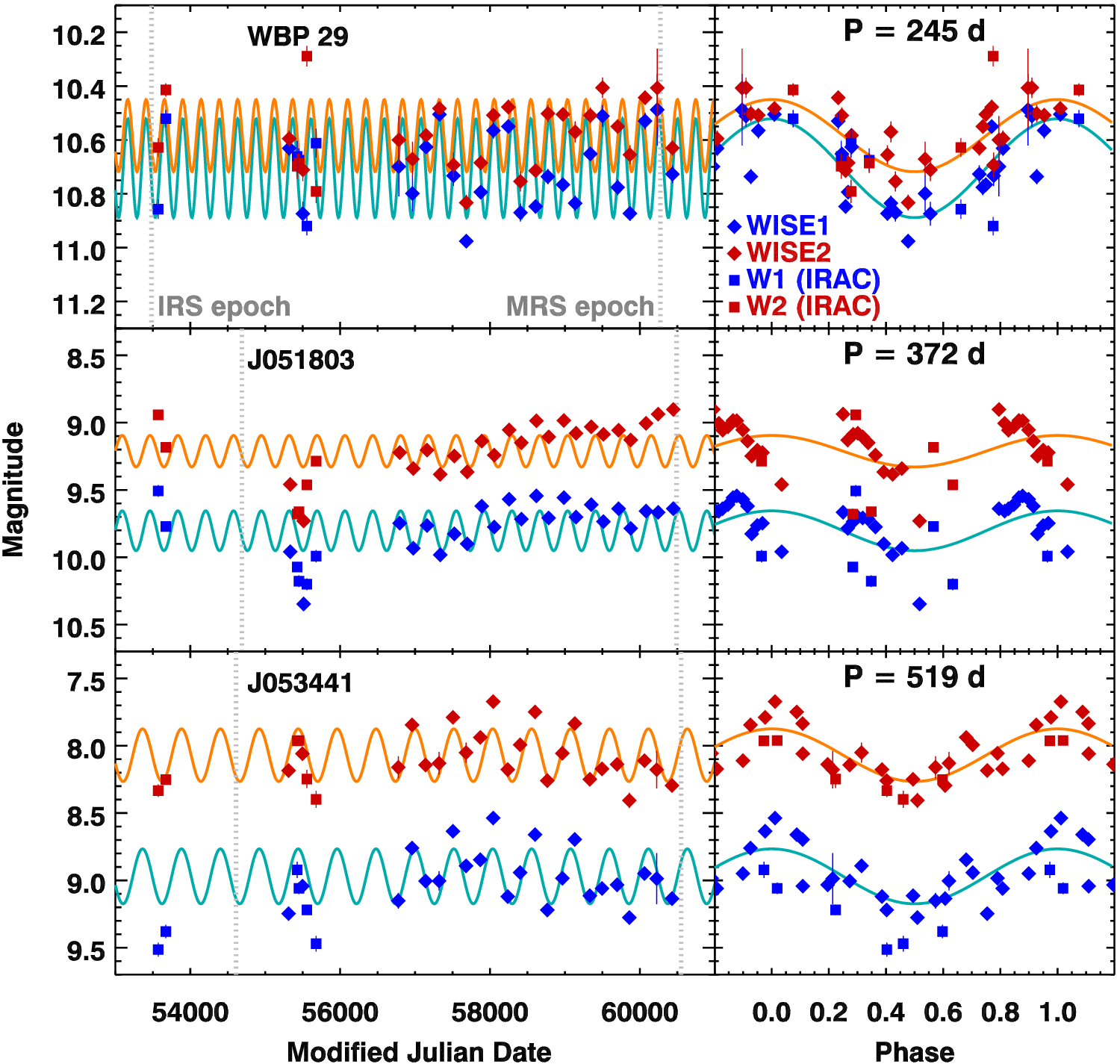}
\caption{Light curves for the three bluest Miras in the sample.  The
figure key is the same as Figure~\ref{f.lc1}.  The top panel has 
half the magnitude range of the bottom two (1.2 mag versus 2.4 mag).  
The nearly 1 yr period of J51803 and the 6 month cadence of the 
WISE data mean that the fitted pulsation amplitudes are not 
trustworthy.
\label{f.lc2}}
\end{figure}

\begin{figure}[!ht] % Fig. 15
\includegraphics[width=240pt]{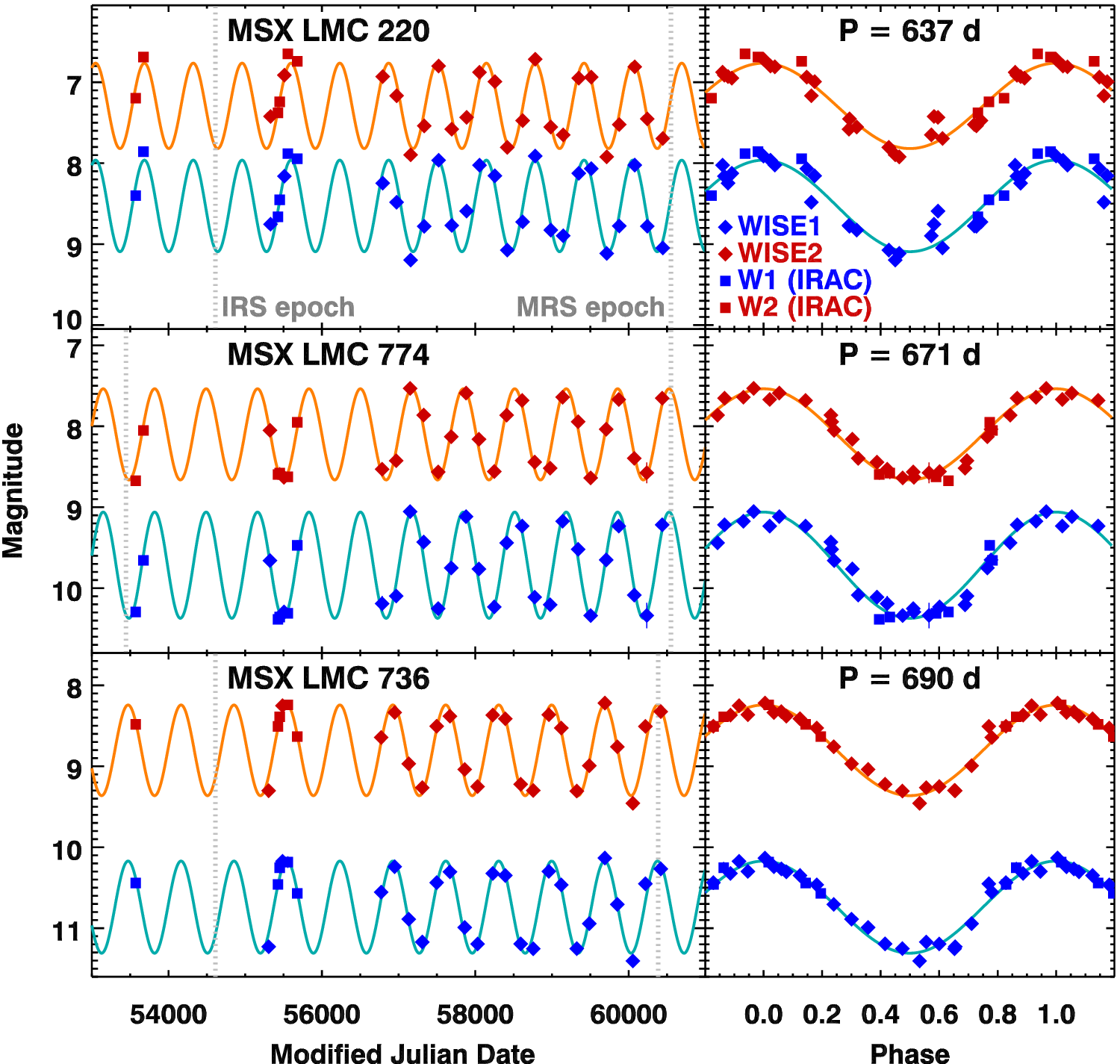}
\caption{Light curves for the three reddest Miras in the sample.  The
figure key is the same as for Figure~\ref{f.lc1}.  All three panels 
have a vertical extent of 4 mag.
\label{f.lc3}}
\end{figure}

\begin{figure*}[!ht] % Fig. 16
\includegraphics[width=480pt]{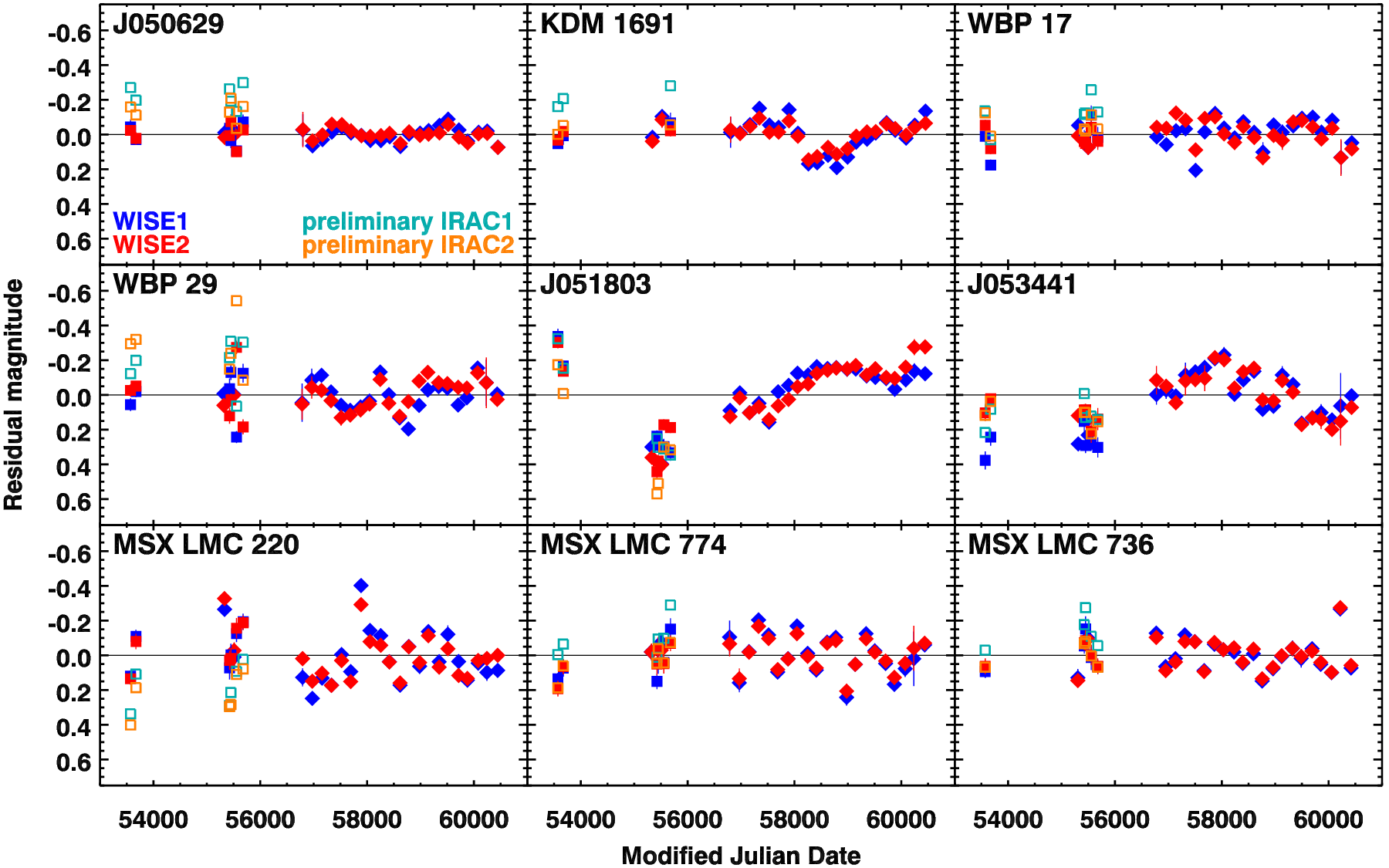}
\caption{Residuals after subtracting the fitted sinusoids from the 
observed light curves, with diamonds for WISE observations and squares 
for color-corrected IRAC data.  The IRAC data after the color corrections 
in Table~\ref{t.corr} are plotted as open squares in light blue (for 
IRAC1) and orange (for IRAC2).  The corresponding data after correcting 
for the residual offsets in Table~\ref{t.lc} are plotted as blue or
red closed squares.
\label{f.res}}
\end{figure*}

Table~\ref{t.lc} gives the periods to nearest day, given the
uncertainties of $\sim$1--5 days.  Figures~\ref{f.lc1} to 
\ref{f.lc3} show the light curves and fitted sinusoids for both 
unphased and phased data, and Figure~\ref{f.res} plots the residuals 
from the fitted sinusoids (data $-$ model).  The following subsections
discuss the results for each star.

\begin{deluxetable*}{lcrrrrcrrcc} % Table 6
\tablecolumns{11}
\tablewidth{0pt}
\tablenum{6}
\tablecaption{Fitted Light Curves}
\label{t.lc}
\tablehead{
  \colhead{ } & \colhead{Pulsation} & 
  \multicolumn{2}{c}{Mean} & \multicolumn{2}{c}{Amplitude\tablenotemark{a,b}} &
  \colhead{Mean} & 
  \multicolumn{2}{c}{IRAC Offsets\tablenotemark{c}} & 
  \multicolumn{2}{c}{Phase When} \\
  \colhead{ } & \colhead{Period\tablenotemark{a}} & 
  \multicolumn{2}{c}{Magnitude} &
  \multicolumn{2}{c}{(mag)}
& \colhead{Zero-phase} & 
  \multicolumn{2}{c}{(mag)} & \multicolumn{2}{c}{Observed by} \\
  \colhead{Target} & \colhead{(d)} & \colhead{W1} & \colhead{W2} &
  \colhead{W1} & \colhead{W2} & \colhead{MJD} & \colhead{W1} & \colhead{W2} &
  \colhead{IRS\tablenotemark{a}} & \colhead{MRS\tablenotemark{a}}
}
\startdata
J050629      & (218 $\pm$ 0.8) & 10.456 & 10.519 &  0.090  &  0.077  &
  53845.5 $\pm$ 3.0 & $-$0.217 & $-$0.135 & (0.93) & (0.48) \\
KDM 1691     &  523 $\pm$ 1.2  &  9.065 &  9.010 &  0.174  &  0.143  &
  53364.4 $\pm$ 2.9 & $-$0.214 & $-$0.034 &  0.46 &  0.74 \\
WBP 17       &  309 $\pm$ 2.0  & 10.400 & 10.319 &  0.269  &  0.216  &
  53608.5 $\pm$ 0.7 & $-$0.146 & $-$0.072 &  0.96 &  0.47 \\
WBP 29       &  245 $\pm$ 0.8  & 10.705 & 10.584 &  0.367  &  0.268  &
  53654.7 $\pm$ 0.9 & $-$0.179 & $-$0.269 &  0.30 &  0.02 \\
J051803      &  372 $\pm$ 4.7  &  9.803 &  9.212 & (0.296) & (0.234) &
  53091.1 $\pm$ 7.0 &    0.013 &    0.128 &  0.29 &  0.89 \\
J053441\tablenotemark{d}
             &  519 $\pm$ 2.9  &  8.970 &  8.070 &  0.408  &  0.391  &
  55434.0 $\pm$ 4.7 & $-$0.161 &    0.015 &  0.40 &  0.85 \\
MSX LMC 220  &  637 $\pm$ 4.5  &  8.528 &  7.292 &  1.131  &  1.055  &
  53685.4 $\pm$ 0.6 &    0.217 &    0.266 &  0.52 &  0.78 \\
MSX LMC 774  &  671 $\pm$ 1.7  &  9.716 &  8.100 &  1.312  &  1.126  &
  53831.5 $\pm$ 0.9 & $-$0.138 & $-$0.005 &  0.43 &  0.01 \\
MSX LMC 736  &  690 $\pm$ 1.4  & 10.740 &  8.803 &  1.138  &  1.120  &
  53472.9 $\pm$ 0.6 & $-$0.123 &    0.005 &  0.64 &  0.02
\enddata
\tablenotetext{a}{Values in parentheses are poorly constrained.}
\tablenotetext{b}{The peak-to-peak amplitude of the fitted sinusoid.}
\tablenotetext{c}{Offsets should be subtracted from the IRAC photometry 
  after the first correction.}
\tablenotetext{d}{The results are from the light curve fitted to just the 
  WISE photometry.}
\end{deluxetable*}

\subsection{J050629}

J050629 is classified as an SRV, and it behaves like one, with a small
pulsation amplitude and no clear pulsation period, resulting in
inconsistent periods in the literature.  The OGLE-III survey reported 
two periods, 741.3 and 154.42 days with similar amplitudes 
\citep{sos09}, while survey for Massive Compact Halo Objects (MACHO)
reported periods of 266.52 and 238.49 days, again with similar 
amplitudes \citep{fra05}.  \cite{slo24} published a light curve with 
a period of 154 days, consistent with the shorter OGLE-III period.

The present analysis includes one more WISE epoch than available to
\cite{slo24}.  Adding it and improving the corrections to the
IRAC photometry result in two possible pulsation periods, 156 or 218
days.  The combined $\chi^2$ residuals for W1 and W2 are smaller for 
the 218 day period, and that is what is reported here.  No one period 
dominates the light curve of this star.  The different published 
periods may well result from what mode, or resonance between modes, 
happened to be strongest at the time each survey obtained its data.  
We can say little about the relative phases of observations for this 
star, but the pulsation amplitudes are small, so that the star was 
likely to have been in a similar physical state for both observations.

\subsection{KDM 1691}

KDM 1691 is an SRV, like J050629, but the literature gives a more 
consistent period of somewhere over 500 d, with reported periods of 
502.77 \citep{sos09}, 518.7 \citep{fra05}, 529 \citep{gro18}, and
531.95 days \citep{kim14}, to take some examples from the literature.
However, both the OGLE-III and MACHO surveys reported weaker overtone 
periods of 285.8 and 261.37 days, respectively, indicating that the 
pulsations are not particularly stable.

Analyzing the WISE data reveals multiple possible periods, at 135, 
170, 197, 279, and 522 days, with minimum $\chi^2$ residuals at 197 
days for W1 and 522 days for W2.  Such complex pulsational behavior 
should be expected for an SRV.  The 522~d period is more consistent 
with the optical results, and because this star is not reddened by 
much dust and the optical data have a much better cadence, it is 
reported here.

The amplitudes of the pulsation cycles in KDM 1691 are twice those of
J050729, but they are still less than 0.2~mag, making this star a weak
pulsator, despite the apparently long pulsation period.  The light 
curve in Figure~\ref{f.lc1} and the residuals in Figure~\ref{f.res} 
show that the pulsation cycle is more complex than a single sinusoid, 
with a notable excursion to fainter magnitudes between MJD 58000 and 
59000 (2017--2020).  All of the above is behavior expected for an SRV.  
Nonetheless, the pulsation period is long enough and characterized 
well enough that we can state confidently that the star was observed 
close to its minimum by the IRS and halfway between minimum
and maximum by the MRS.

\subsection{WBP 17}

WBP 17 is the third SRV in the sample, and its period is consistently
close to 310 days in the literature.  The OGLE-III survey reported 
314.3 days \citep{sos09} and the MACHO survey reported 308.45 days 
\citep{fra05}.  \cite{ou22} reanalyzed the OGLE-III data and found a 
period of 315.44 days.  Our analysis suggests that pulsation periods of 
244 and 309 days fit the light curves with similar residuals, but the 
optical data are more finely sampled.  Consequently, we report the 
period closest to $\sim$310idays.

The pulsation amplitudes of WBP 17 are the strongest of the three SRVs 
in the sample, but evidence for some irregularities in its behavior can 
still be seen in its light curve and its residuals at a level of 
$\sim$0.2 mag.  Despite those deviations from a sinusoid, the light 
curve is clean enough to show that the IRS observation came close to 
maximum and the MRS observation close to minimum.

\subsection{WBP 29}

WBP 29 is the least reddened of the six Miras in the sample, and it 
also has the shortest pulsation period, with OGLE-III reporting 246.7 
days \citep{sos09} and MACHO reporting 245.82 days \citep{fra05}.  
Seven other investigators found periods between 240 and 247 days 
\citep{gro04, ita04, spa11, kim14, gro18, iwa21}.  Our analysis of the 
IRAC and WISE data results in a period of 245 days.  The fitted light 
curves imply that the IRS observation was obtained approaching minimum 
and the MRS observation at maximum.  

\subsection{J051803}

Previous studies indicate a pulsation period for J051803 close to 1 
yr.  \cite{sos09} found that the strongest pulsations had a 348.9 day
period, \cite{gro18} reported 352 days, and \citep{iwa21} reported 
362.44 days.  A period close to 1 yr is particularly difficult for 
the WISE cadence, but the IRAC epochs may have broken the degeneracy.  
For the first pass, instead of using WISE data to find a period, we 
adopted the period of 362.44 day and used that to determine the 
corrections for the IRAC photometry.  

The light-curve residuals for J051803 in Figure~\ref{f.res} indicate
that the star has long-term variations, with a peak-to-peak amplitude 
of $\sim$0.5~mag.  To avoid the impact of this long-term trend, we 
determined the corrections to the IRAC data using the four SAGE-Var
epochs and the two WISE epochs in the interval 55000 $<$ MJD $<$ 
56000.  Combining the corrected IRAC photometry with WISE gives a 
pulsation period of 372 days.  This pulsation period is still close to 
a year, and the full pulsation cycle is poorly sampled by the IRAC and 
WISE photometry.  For that reason, the pulsation amplitude is poorly
constrained, as indicated in Table~\ref{t.lc}.

The fitted periods for J051803 have been increasing with time, but
whether this result is meaningful or not is an open question.  The 
zero-phase epoch for the fitted sinusoids differs by 14 days between 
W1 and W2, the largest difference for the stars in our sample.  That
might be a result of a period close to 1 yr, or it could arise from 
the long-term variations.  The residuals show a hint of periodicity, 
but the available data do not constrain a possible period well.  
Fitting both the WISE data and the corrected IRAC data yields a 
period of $\sim$6140 days in W1 and $\sim$7130 days in W2, which is
a rather significant difference of $\sim$15\%.  While the period may
not be well constrained, the peak-to-peak amplitude of a sinusoid
fitted to these residuals is significant:  0.93 mag in W1 and 1.0 mag
in W2.  These long-period results should be viewed more as 
suggestions than meaningful measurements.  With all of these 
uncertainties, the only safe conclusion is that J051803 has a complex 
light curve.

The fitted light curve for J051803 suggests that the IRS epoch was
obtained as the source was roughly halfway from maximum to
minimum and the MRS epoch was approaching maximum.  But 
all the caveats for the light curve for this source should be kept in 
mind.

\subsection{J053441}

J053441 has a light curve similar to J051803, except that the primary
pulsation period is considerably longer than one year.  Previous
investigations found periods between 489 and 548 days \citep{sos09,
spa11, gro20, iwa21, ou22}.  Our analysis of the WISE photometry 
results in a period of 519 days.  As with J051803, the light curve 
shows long-term variations (see Figure~\ref{f.res}), forcing us to 
determine a correction for the IRAC data using just the data with 
55000 $<$ MJD $<$ 56000.  Despite those corrections, adding the IRAC 
data did not help us determine a primary period, because the 
curve-fitting software failed to converge at the minimum $\chi^2$ 
residuals in W2.  The period in W1 was 522 days, and that in W2 at 
least 523 days.  Table~\ref{t.lc} gives the results for just the WISE 
data, although the uncertainty in the period reflects the range in the 
fitted period introduced by adding the IRAC data.

J053441, like J051803, shows possible periodicity in its residuals.
Fitting both the WISE data and the corrected IRAC data gives periods
of $\sim$7700 days for W1 and $\sim$5700 days for W2, but the 
discrepancy between the two filters is even larger than for J051803.  
The peak-to-peak amplitudes are smaller:  0.90 mag in W1 and 0.57 mag 
in W2.  The long-term variations are clear enough, but the available 
data do not allow them to be properly quantified.

\subsection{MSX LMC 220}

For MSX~LMC~220, we find a period of 644 days with just the WISE data.
Adding the corrected IRAC data shifts this result to 637 days.  
Previous results have similar periods of 625 and 632 days 
\citep{gro09, gro18}, although OGLE-III reported periods of 313 and 
1805 days \citep{sos09}.  The light curves based on the IRAC and WISE 
data indicate fairly steady pulsations over the past 20 yr in the 
3--5~\mum\ range (Figure~\ref{f.lc2}), but the residuals in 
Figure~\ref{f.res} show deviations of up to $\sim$0.4~mag in some 
isolated epochs, suggesting that the pulsations are not completely 
settled.  The pulsation cycle is long 
enough and the pulsations steady enough to conclude with some confidence 
that the IRS epoch came close to minimum, while the MRS epoch came as 
MSX~LMC~220 was roughly halfway to maximum.

\subsection{MSX LMC 774}

Moving to the red end of the Mira sequence leads to steadier and
stronger pulsations, and MSX~LMC~774 follows that trend.  Its dust 
shell effectively erases the star from the optical skies, so it does 
not appear in the OGLE-III and MACHO surveys.  Using the IRAC data and
the limited number of WISE epochs available at the time, \cite{slo16}
found a period of 670 days, and with more epochs and a slightly 
different technique, \cite{gro22} found periods of 680 days in W1 and 
678 days in W2.  Using even more WISE epochs, we find a period of 668 
days.  The well-behaved pulsation cycle and long period of MSX~LMC~774 
constrain the epochs of the IRS and MRS spectroscopy well, with the 
IRS observing the star just before minimum and the MRS very close to 
maximum.

\subsection{MSX LMC 736}

MSX LMC 736 is another Mira too reddened by its own dust to appear in
optical variability surveys.  Previous measurements of its period have
been based on the six SAGE and SAGE-Var epochs and increasing numbers 
of epochs from WISE:  686 \citep{slo16}, 683 \citep{gro18}, 672 
\citep{gro20}, and 690 days \citep{slo24}.  This work duplicates the 
last period, 690 days, even though it adds one WISE epoch and corrects 
the IRAC photometry for residuals.  As with MSX~LMC~774, the light 
curve constrains the IRS and MRS epochs well, with the first just 
after minimum and the second right at maximum.

\section{Luminosities \label{s.lum}} % Appendix B

\begin{deluxetable*}{lccccc} % Table 7
\tablecolumns{6}
\tablewidth{0pt}
\tablenum{7}
\tablecaption{Bolometric Luminosities and Magnitudes}
\label{t.lum}
\tablehead{
  \colhead{Target} & \multicolumn{4}{c}{Bolometric Luminosity (L$_{\odot}$)} &
  \colhead{$M_{\rm{bol}}$ (mag)} \\
  \colhead{ } & \colhead{Groenewegen} & \colhead{PySSED} & 
  \colhead{Curated SED} & \colhead{Mean\tablenotemark{b}} & \colhead{ } \\
  \colhead{ } & \colhead{\& Sloan (2018)\tablenotemark{a}} & 
  \colhead{Method} & \colhead{and Spectra} & \colhead{ } & \colhead{ }
}
\startdata
J050629     &  (3489)  &  3818  &  3824  &   3821 $\pm$   3 & 
  $-$4.21 $\pm$ 0.00 \\
KDM 1691    &  12,707  & 13,595 & 14,653 & 13,652 $\pm$ 562 & 
  $-$5.59 $\pm$ 0.04 \\
WBP 17      &  (7268)  &  5201  &  5824  &   5513 $\pm$ 312 & 
  $-$4.60 $\pm$ 0.06 \\
WBP 29      &   5032   &  5084  &  4910  &   5009 $\pm$  52 & 
  $-$4.50 $\pm$ 0.01 \\
J051803     &  (4851)  &  2968  &  3570  &   3269 $\pm$ 301 & 
  $-$4.04 $\pm$ 0.10 \\
J053441     &   8198   & 10,232 &  9145  &   9192 $\pm$ 588 & 
  $-$5.16 $\pm$ 0.07 \\
MSX LMC 220 & (23,388) & 16,290 & 15,230 & 15,760 $\pm$ 530 & 
  $-$5.74 $\pm$ 0.04 \\
MSX LMC 774 &   7574   &  7342  &  9620  &   8179 $\pm$ 723 & 
  $-$5.03 $\pm$ 0.10 \\
MSX LMC 736 &   9311   &  7168  &  8150  &   8210 $\pm$ 619 & 
  $-$5.04 $\pm$ 0.08
\enddata
\tablenotetext{a}{Values not used in mean luminosity are in parentheses.}
\tablenotetext{b}{The uncertainties are the uncertainty in the mean.}
\end{deluxetable*}

Table~\ref{t.lum} presents bolometric luminosities for the stars in 
our sample, determined using two methods and compared to the results 
from \cite{gro18}.  Our first method is to apply the PySSED software
(version 1.1), which constructs a spectral energy distribution (SED) 
for a source from an automated lookup of the available photometry in 
VizieR and integrates the result through wavelength space 
\citep{mcd24, mcd25}.  Our second method of determining the bolometric 
luminosity uses a curated set of photometry and the spectra from the 
IRS and MRS.  

For the second (curated) method, the mid-infrared photometric data
were limited to the mean 24~\mum\ photometry from the SAGE survey 
\citep{mei06} and the mean results in the WISE filters at 3.4 and 
4.6~\mum\ reported in Appendix~\ref{s.lc} (including the 
color-corrected IRAC photometry from SAGE and SAGE-Var).  We also used 
near-infrared photometry from the Two Micron All Sky Survey 
\citep[2MASS;][]{skr06}, 2MASS~6X (same reference), photometry from 
the Infrared Spectral Survey \citep{kat07}, the Deep Near Infrared 
Survey of the Southern Sky \citep[DENIS;][]{cio00}, and the Vista 
Magellanic Cloud (VMC) survey DR6 \citep{cio11}.  Photometry at 
shorter wavelengths came from the OGLE-III survey \citep{sos09}, 
the MACHO survey \citep{fra08}, the Magellanic Clouds Photometric 
Survey \citep[MCPS;][]{zar04}, the SkyMapper Southern Sky Survey 
DR4 \citep{onk24}, and Gaia DR3 \citep{gaia23}.  We did not include 
the K$_{\rm s}$ photometry from the VMC because it gave the faintest 
magnitude at K$_{\rm s}$ in five of the nine sources.  We omitted the 
MACHO R$_c$ magnitudes, because they were low outliers in all four 
sources for which data were available.  We also omitted the MCPS 
i-band data for J053441 and MSX~LMC~220 because they were outliers and 
one of the DENIS I-band magnitudes for MSX~LMC~220 for the same reason.  
For both J051803 and MSX~LMC~736, some care was required to avoid 
neighboring sources.  We averaged the remaining data at I 
(0.77--0.80~/um), J (1.22--1.25~\mum), H (1.61--1.67~\mum), and 
K$_{\rm s}$ (2.23--2.16~\mum).

\begin{figure}[!ht] % Fig. 17
\includegraphics[width=240pt]{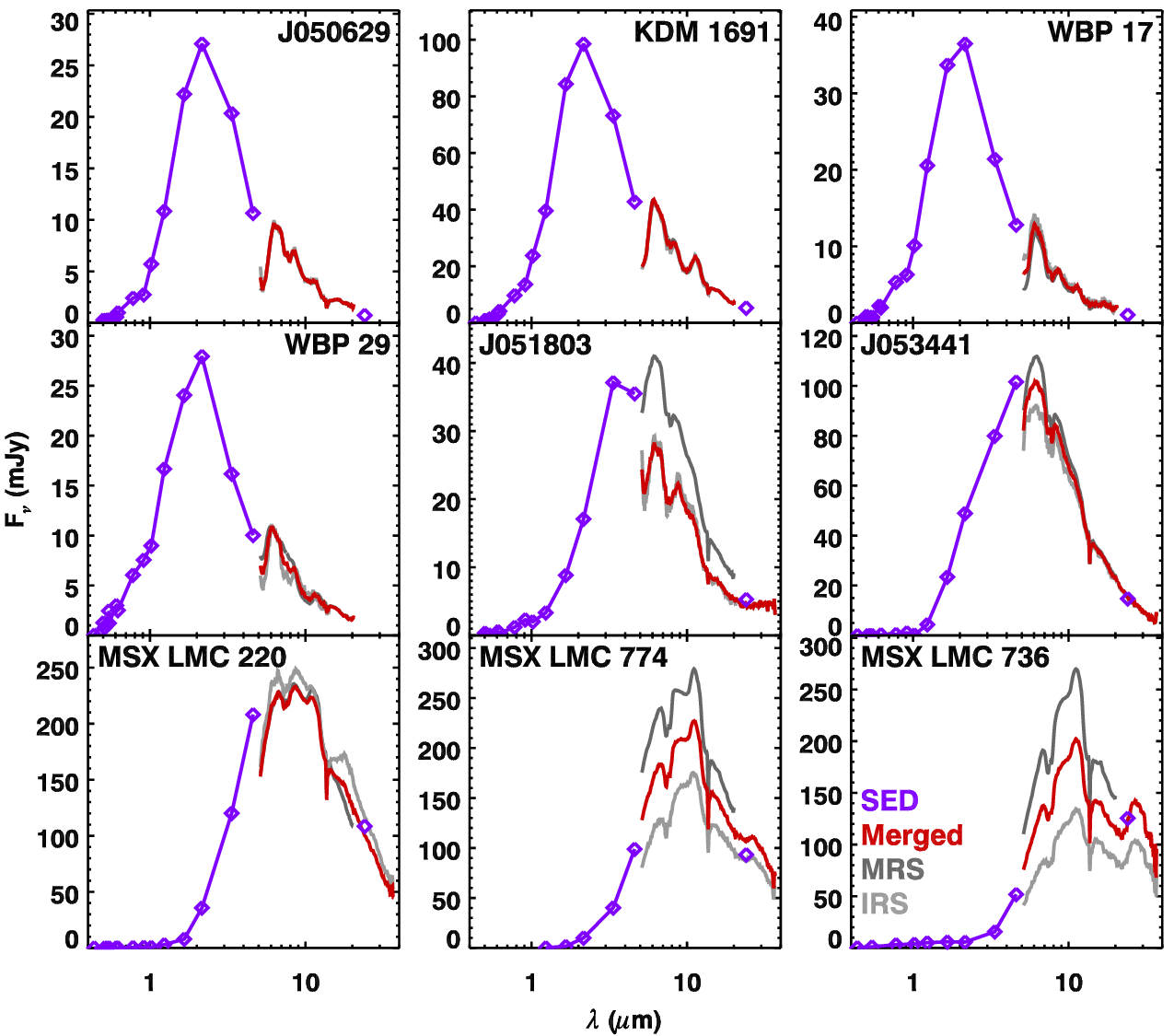}
\caption{SEDs and merged spectra from the MRS and IRS for the nine 
stars in the sample.  The plotted data are not corrected for 
interstellar absorption.
\label{f.sed}}
\end{figure}

We combined the MRS and IRS data by regridding the MRS to the IRS,
truncating the MRS beyond 20~\mum, averaging the spectra where
they overlapped, and normalizing the remaining spectral data to
align with the average.  To normalize the resulting spectra to the
approximate mean of the light curve, we used the phase information for 
the epochs in Appendix~\ref{s.lc}.  For six of the nine carbon stars, 
we forced the combined spectrum to the average of the two.  The 
exceptions were KDM~1691 and MSX~LMC~220, where we forced the IRS to 
align with the MRS, and J051803, where we did the opposite.  
Figure~\ref{f.sed} shows the resulting SEDs for all nine sources, along 
with the spectra from the MRS and IRS.

The PySSED code uses Gaia-based extinction maps from \cite{lal22}.  
\cite{gro18} assumed $A_v$ = 0.15 in the LMC, and we followed their 
lead.  We used the interstellar extinction from \cite{rie85} to 
1.35~\mum\ and that from \cite{chi06} past 1.35~\mum.  The extinction 
affected the results by $\sim$3\% for the bluer sources and 
$\sim$1\%--2\% for the redder sources, meaning that the specific 
choice for the extinction correction has little effect.

We integrated the combination of SED and spectra to determine the 
bolometric luminosity of each source, using a Wien tail for a 3000 K 
blackbody to extrapolate to the blue from the shortest observed 
wavelength and a Rayleigh-Jeans tail to extrapolate from the longest 
observed wavelength to the red.  For the three SRVs and WBP~29, the 
usable spectral data only extended as far as 20~\mum, so we included 
the MIPS 24~\mum\ photometry in the SED.  For the remaining five 
Miras, the spectral data extended to 38~\mum\ and we omitted the MIPS 
photometry.

For the luminosities and bolometric magnitudes, we assumed a distance
modulus to the LMC of 18.48 mag \citep{pie19}.  We shifted the
results from \cite{gro18} to this distance; they assumed a distance of 
50.0 kpc ($m-M$ = 18.495 mag).

As Table~\ref{t.lum} shows, the PySSED method generally obtained similar 
results to the combination of curated SEDs and MRS and IRS spectra, in no 
small part because both methods shared much of the same photometry.  
Given the work involved in curating the photometry, the similar results
are a strong argument for using the more automated PySSED method.  
\cite{gro18} used the photometry available at that time to construct a 
model of the star and from that determined the bolometric luminosity.  
That method produced results that differed more significantly.  We 
excluded the model-based luminosities when they doubled the standard 
deviation estimated from our two methods (Table~\ref{t.lum} shows the 
excluded values in parentheses).  The remaining five luminosities were 
averaged with the results from our two methods to provide the mean 
luminosities in Table~\ref{t.lum} (and Table~\ref{t.sample}).

\section{Fitting the band structure in the spectra 
\label{s.flat}} % Appendix C

\begin{figure}[!ht] % Fig. 18
\includegraphics[width=240pt]{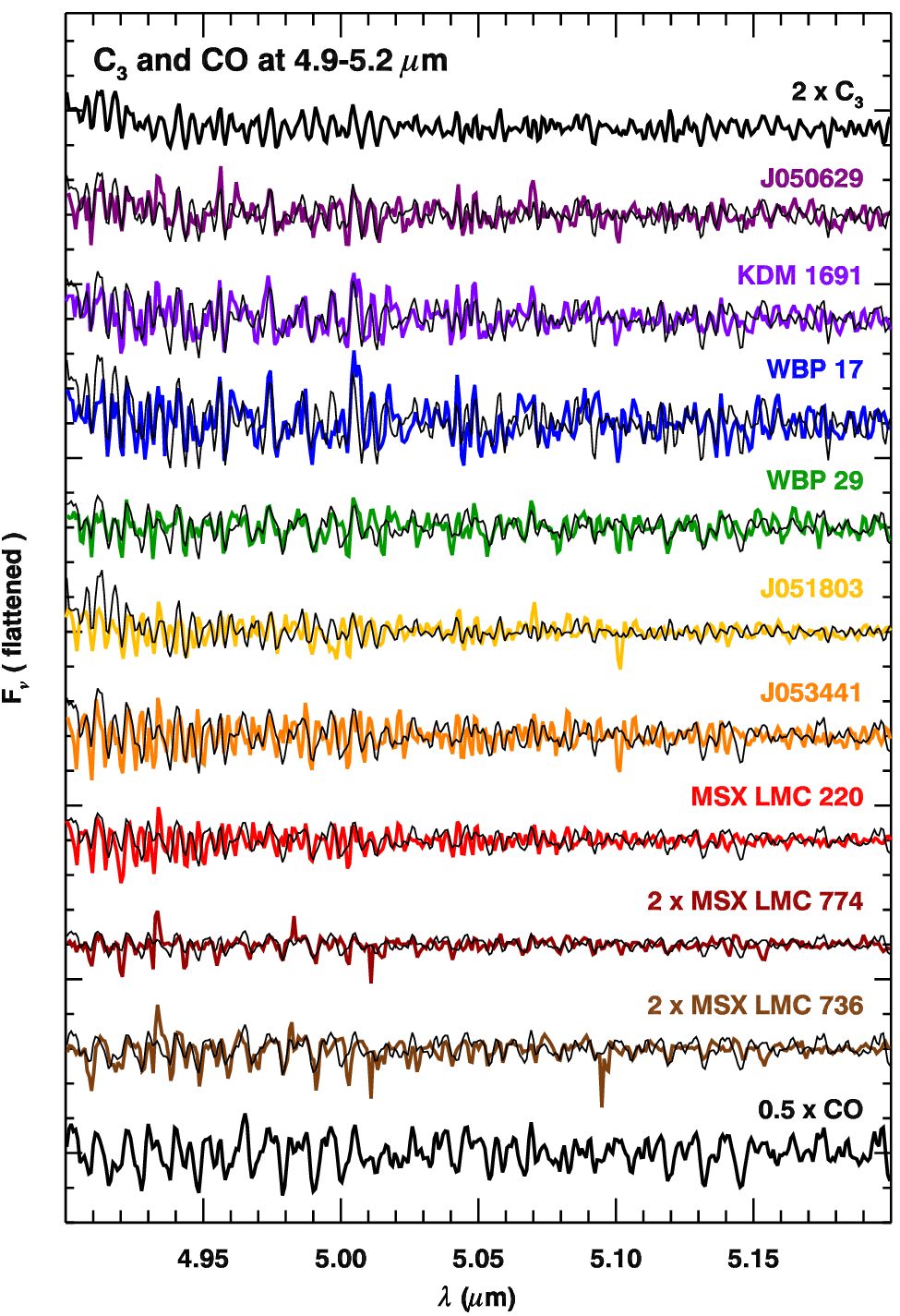}
\caption{The flattened spectra at 4.9--5.2~\mum\ for the nine 
carbon stars in the sample (in color) and the best-fitting combination 
of molecular spectra for each (black).  The C$_3$ spectrum at the 
top is from the synthetic spectrum based on an effective temperature 
of 3100 K.  The CO spectrum at the bottom is for 2800 K.  Both are for 
a C/O ratio of 2.0.  The observed spectra are shifted to correct for 
the radial velocities in Table~\ref{t.vel}.  This figure will be part
of a figure set of 8 images in the published paper.
\label{f.flat05a}}
\end{figure}

The figure set of Figure~\ref{f.flat05a} compares the flattened 
spectra of the carbon stars in our sample to the best-fitting 
combination of absorbing molecules, using synthetic spectra generated 
from a 2800 K model for the SRVs and from 3100 K for the Mira 
variables.  The plotted spectra cover the eight wavelength intervals 
defined in Section~\ref{s.detailed}.  The observed spectra are 
corrected for the radial velocities determined in Section~\ref{s.vel} 
and listed in Table~\ref{t.vel}.  

\begin{figure}[!ht] % Fig. 19
\includegraphics[width=240pt]{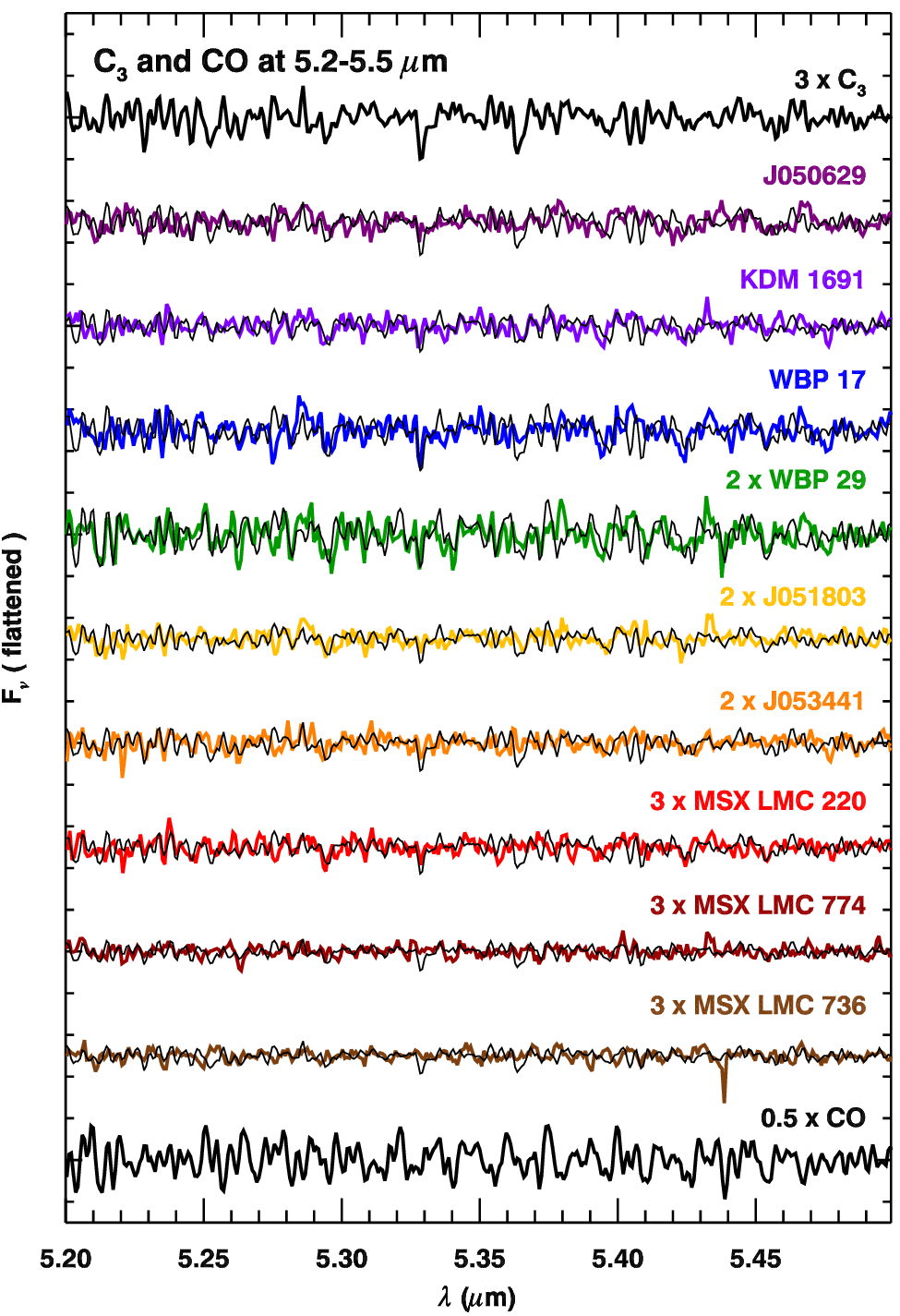}
\caption{As Figure~\ref{f.flat05a}, but for 5.2--5.5~\mum.
This figure will be part of Figure Set 18 in the published paper.
\label{f.flat05b}}
\end{figure}

\begin{figure}[!ht] % Fig. 20
\includegraphics[width=240pt]{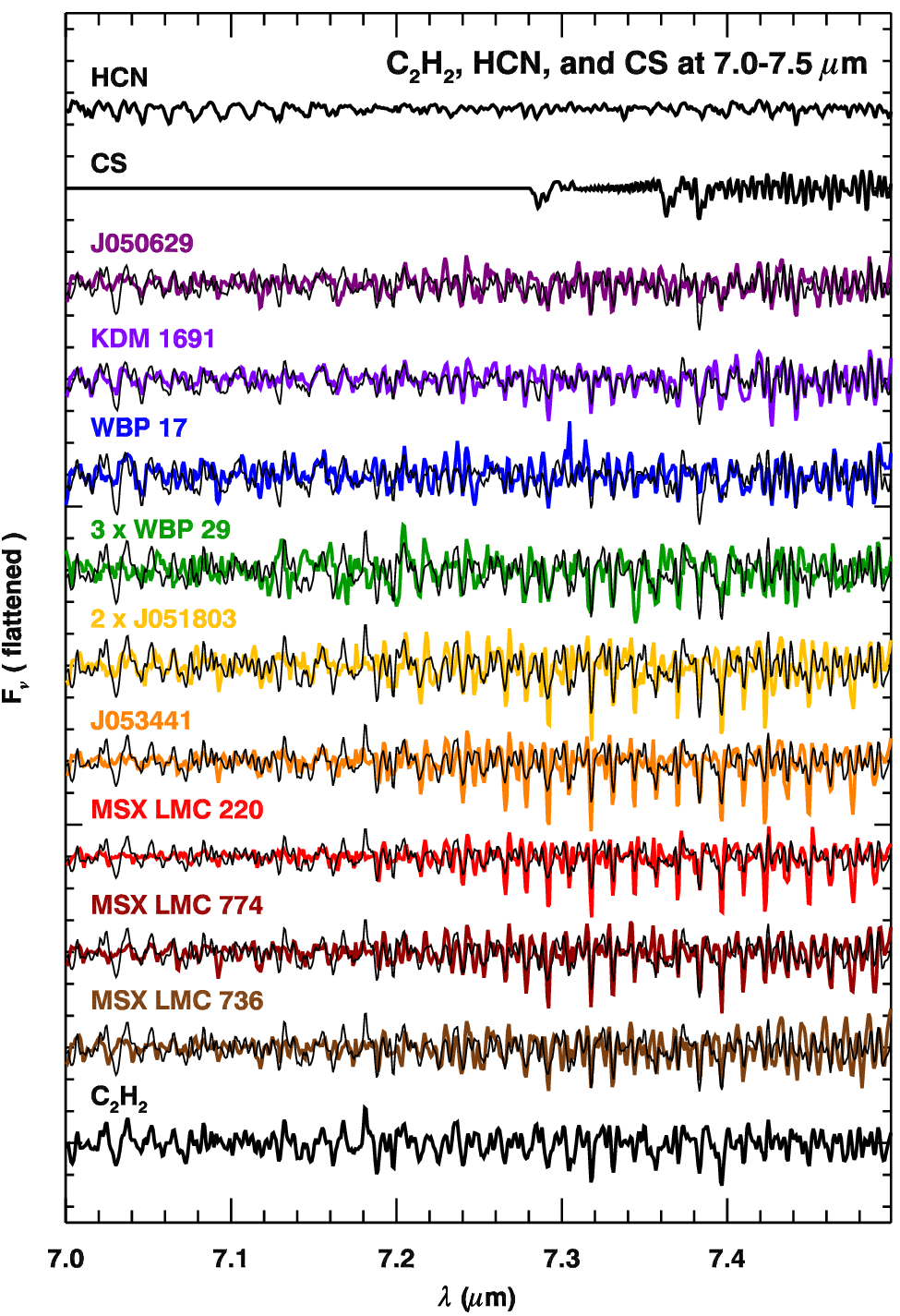}
\caption{The flattened spectra at 7.0--7.5~\mum, with the observed
spectra in color and the fitted molecular spectra in black.  The HCN
and CS at the top are from the synthetic spectra with $T_{\rm eff}$ =
3100 K, and the C$_2$H$_2$ at the bottom is from the 2800 K spectrum.
Both have a C/O ratio of 2.0.  The observed spectra are corrected for
the radial velocities in Table~\ref{t.vel}.
This figure will be part of Figure Set 18 in the published paper.
\label{f.flat07a}}
\end{figure}

\begin{figure}[!ht] % Fig. 21
\includegraphics[width=240pt]{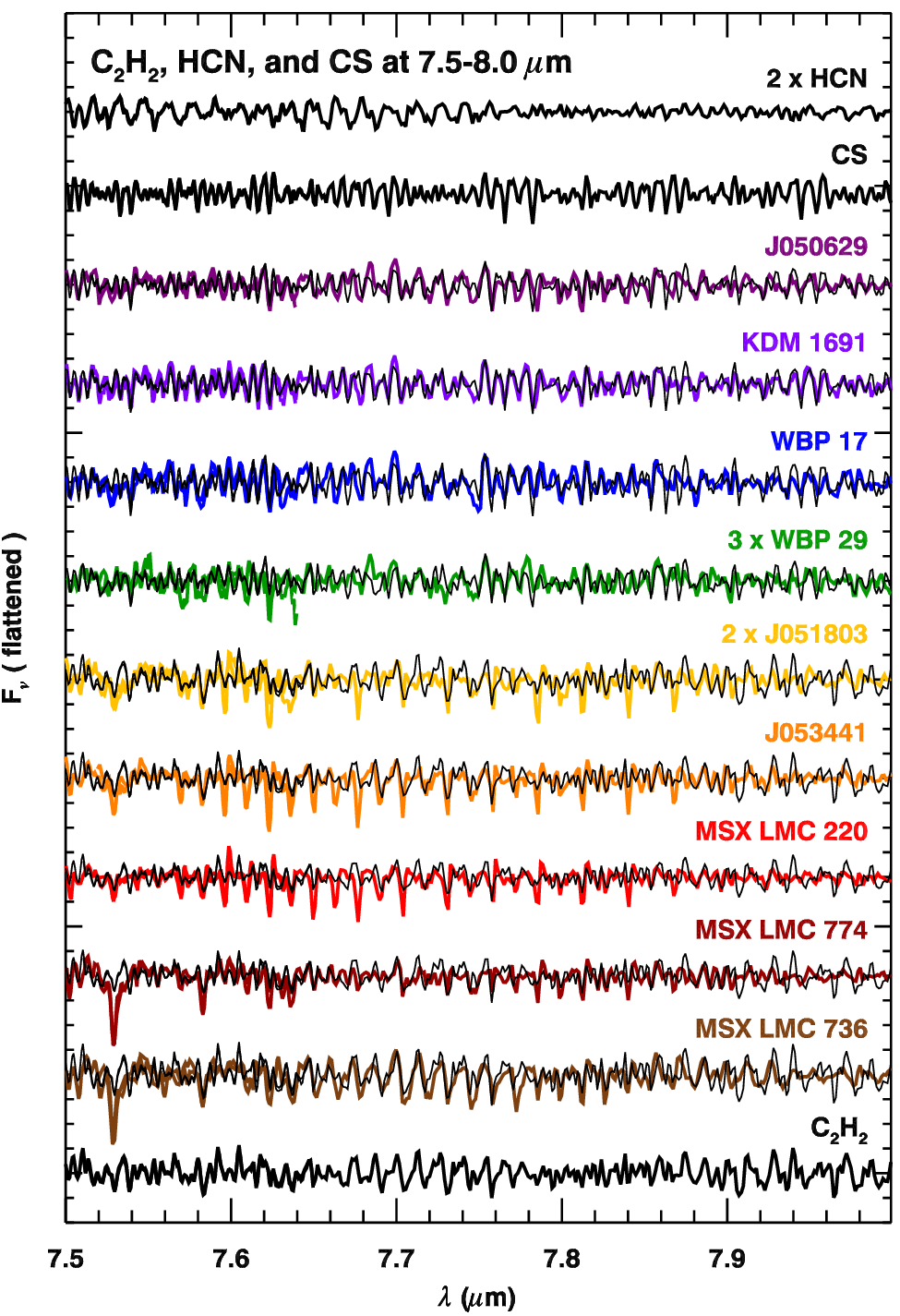}
\caption{As Figure~\ref{f.flat07a}, except for 7.5--8.0~\mum.
\label{f.flat07b}}
\end{figure}

\begin{figure}[!ht] % Fig. 22
\includegraphics[width=240pt]{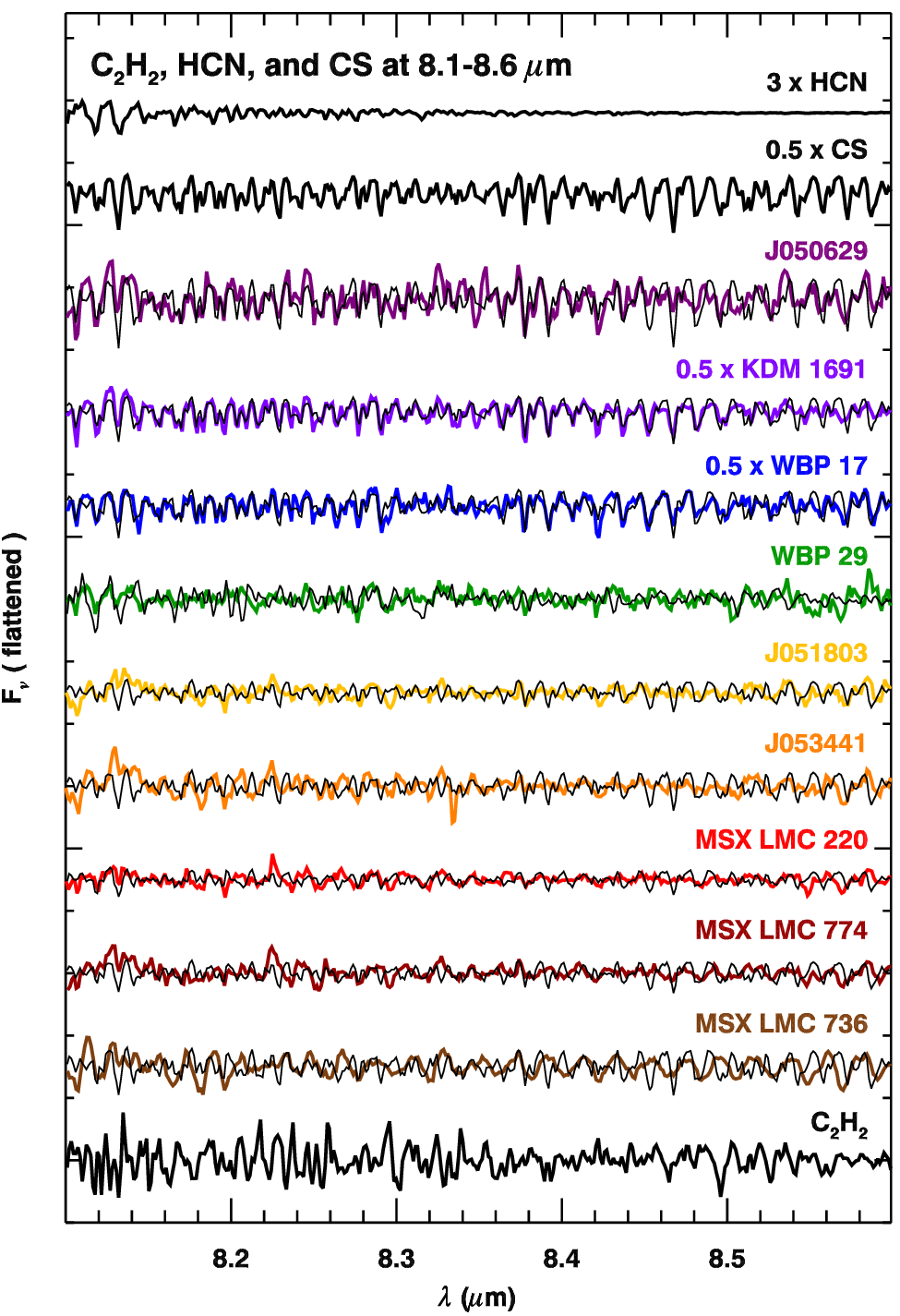}
\caption{As Figure~\ref{f.flat07a}, but for the interval 8.1--8.6~\mum.
\label{f.flat08}}
\end{figure}

\begin{figure}[!ht] % Fig. 23
\includegraphics[width=240pt]{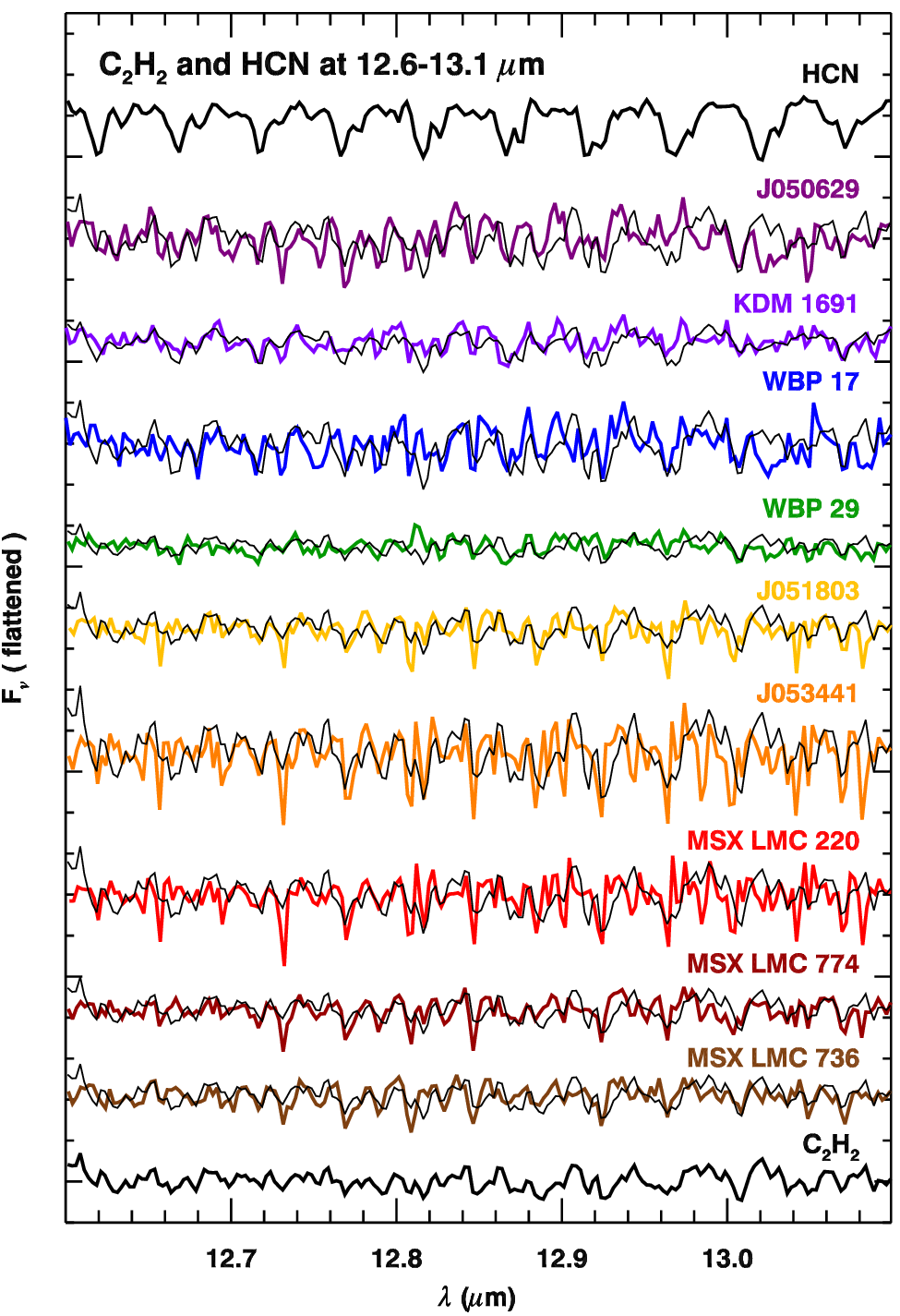}
\caption{The flattened observed spectra at 12.6--13.1~\mum\ (in
color) compared to fitted molecular spectra (black).  The
HCN spectrum at the top is for $T_{\rm eff}$ = 3100 K and a
C/O ratio = 2.0.  The C$_2$H$_2$ at the bottom is for 2800 K
and the same C/O ratio.  The observed spectra are corrected for
the radial velocities in Table~\ref{t.vel}.  This figure will be
part of Figure Set 18 in the published paper.
\label{f.flat12}}
\end{figure}

\begin{figure}[!ht] % Fig. 24
\includegraphics[width=240pt]{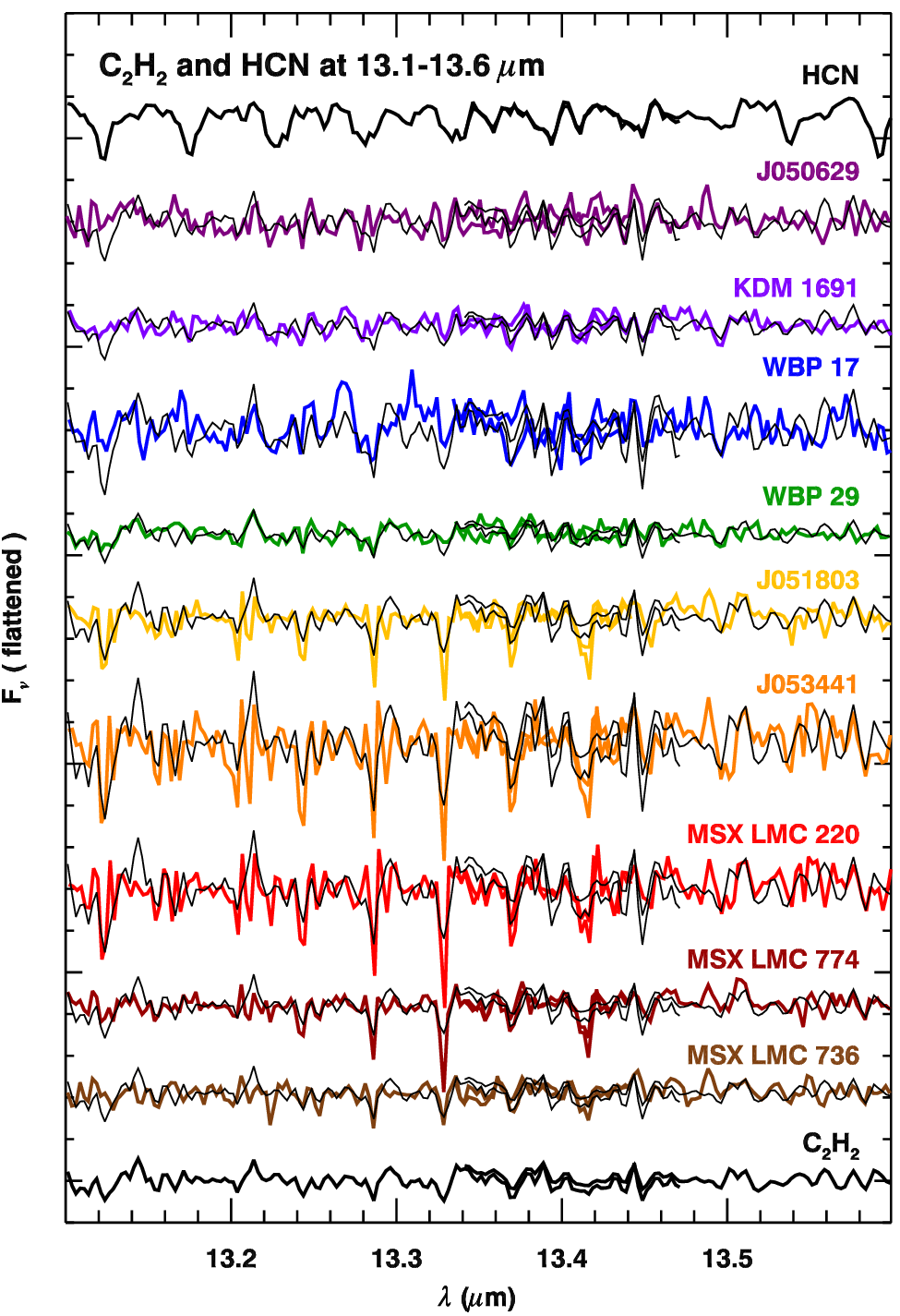}
\caption{As Figure~\ref{f.flat12}, but for 13.1--13.6~\mum.
\label{f.flat13}}
\end{figure}

\begin{figure}[!ht] % Fig. 25
\includegraphics[width=240pt]{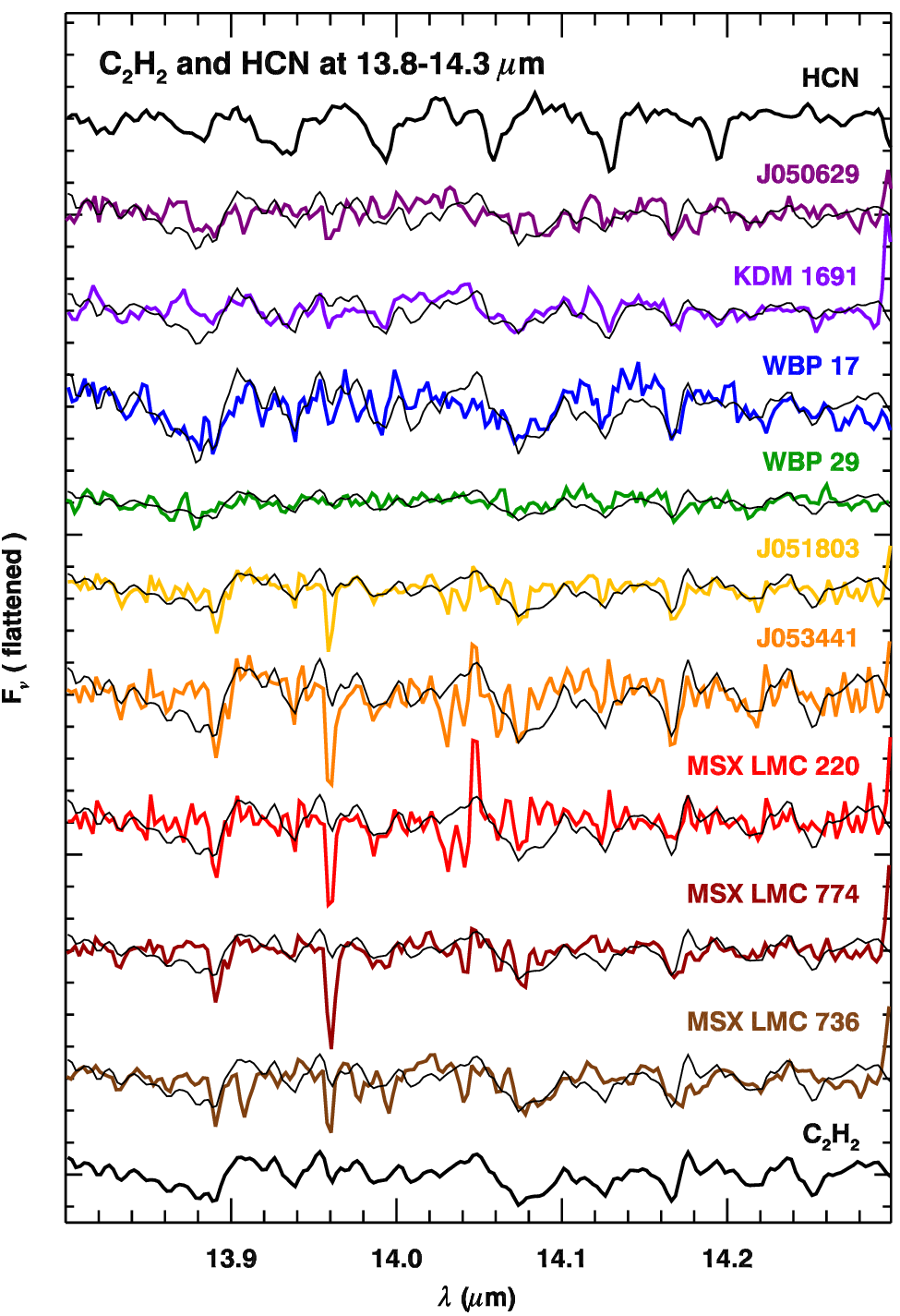}
\caption{As Figure~\ref{f.flat12}, but for 13.8--14.3~\mum.
\label{f.flat14}}
\end{figure}
\section{Multiepoch Spectra from ISO\label{s.sws}} % Appendix D

\begin{deluxetable*}{lrrccllc} % Table 8
\tablecolumns{8}
\tablewidth{0pt}
\tablenum{8}
\tablecaption{Comparison Carbon Stars in the Galaxy}
\label{t.galc}
\tablehead{
  \colhead{Target} & \colhead{R.A.} & \colhead{Decl.} &
  \colhead{Spec.} & \colhead{Spec.} & 
  \multicolumn{2}{c}{Variability\tablenotemark{c}} & 
  \colhead{[6.4]$-$[9.3]\tablenotemark{d}} \\
  \colhead{ } & \colhead{ } & \colhead{ } & \colhead{Type} & 
  \colhead{Ref.\tablenotemark{b}} & \colhead{Type} & \colhead{Period} &
  \colhead{ } \\
  \colhead{ } & \multicolumn{2}{c}{(J2000)\tablenotemark{a}} &
  \colhead{ } & \colhead{ } & \colhead{ } & \colhead{(days)} & \colhead{(mag)}
}
\startdata
R Scl    &  21.783725 & $-$32.801921 & C6.4p & S44 & SRb  & 370    &  0.27 \\
V CrB    & 237.380467 & $+$39.571637 & C6,2e & S44 & Mira & 357.63 &  0.40 \\
V Cyg    & 310.326115 & $+$48.141336 & C5,3e & Y75 & Mira & 421.27 &  0.52 \\
S Cep    & 323.803430 & $+$78.624496 & C6.3e & Y75 & Mira & 484.4  &  0.30 
\enddata
\tablenotetext{a}{From Gaia EDR3 \citep{gaia21}.}
\tablenotetext{b}{S44 = \cite{san44}; Y75 = \cite{yam75}.}
\tablenotetext{c}{\cite{sam17}.}
\tablenotetext{d}{\cite{lei08}.}
\end{deluxetable*}

\begin{deluxetable}{lll} % Table 9
\tablecolumns{3}
\tablewidth{0pt}
\tablenum{9}
\tablecaption{SWS Observing Log}
\label{t.sws}
\tablehead{
  \colhead{Target} & \colhead{TDT\tablenotemark{a}} & \colhead{Obs.\ Date} \\
  \colhead{ } & \colhead{ } & \colhead{(MJD)}
}
\startdata
R Scl   & 24701012 & 50285 \\
        & 37801213 & 50414 \\
        & 37801443 & 50414 \\
        & 39901911 & 50436 \\
        & 41401514 & 50451 \\
        & 56900115 & 50606 \\
V CrB   & 11105149 & 50149 \\
        & 25502252 & 50293 \\
        & 42200213 & 50459 \\
        & 42300201 & 50460 \\
        & 47600302 & 50513 \\
        & 57401003 & 50611 \\
        & 67600104 & 50712 \\
V Cyg   & 08001855 & 50118 \\
        & 42100111 & 50458 \\
        & 42300307 & 50460 \\
        & 51401308 & 50551 \\
        & 59501909 & 50632 \\
        & 69500110 & 50731 \\
S Cep   & 56200926 & 50599 \\
        & 75100424 & 50787
\enddata
\tablenotetext{a}{Time Designated Target number.  The first three digits
give the approximate number of days since the launch of ISO.}
\end{deluxetable}

\begin{figure}[!ht] % Fig. 26
\includegraphics[width=240pt]{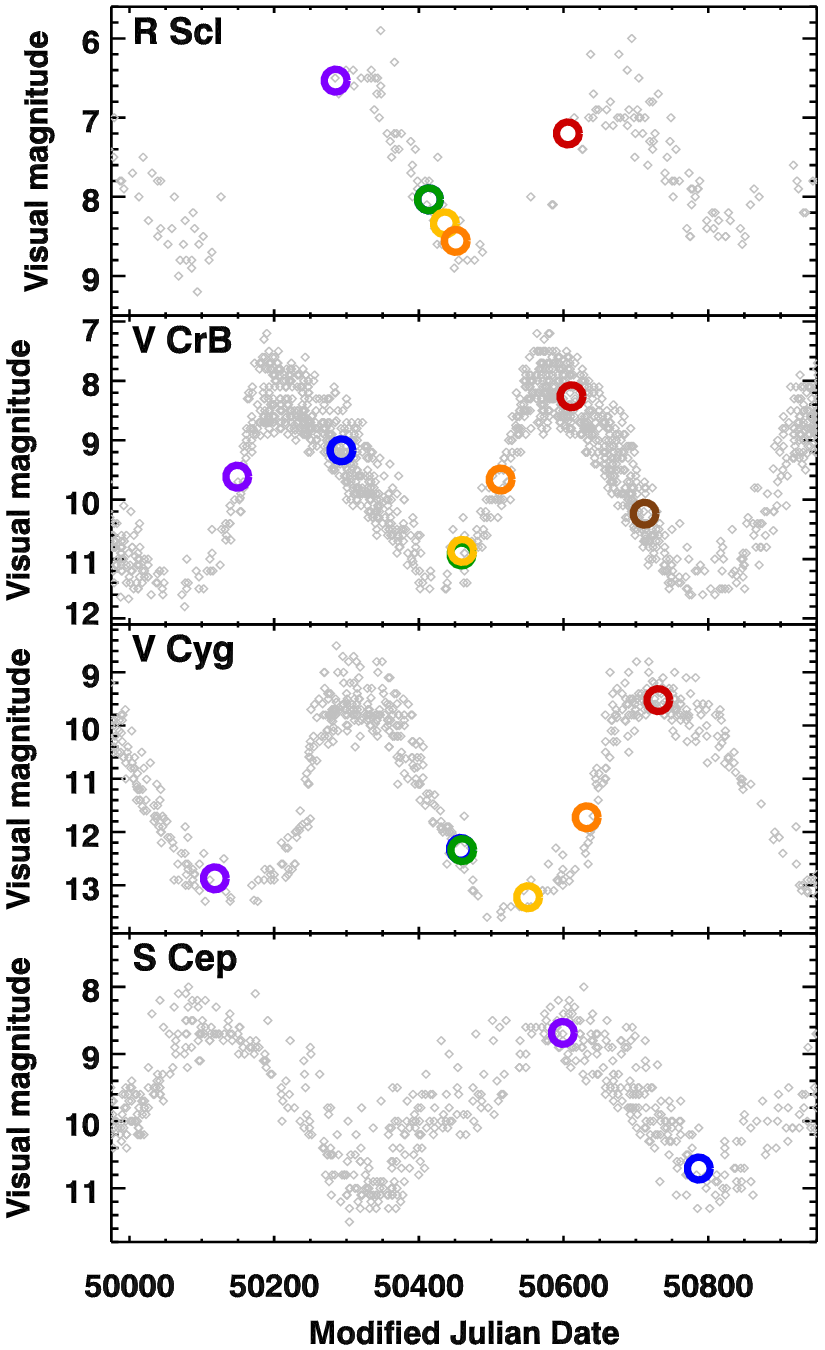}
\caption{Light curves from the AAVSO in gray with the epochs of the 
SWS observations plotted in color.  The magnitudes at the SWS epochs 
are estimated from AAVSO observations with $\pm$20 days.  In the 
panels for R~Scl and V~Cyg, the second and third spectra were obtained 
the same day, so the green plotting symbol completely covers the blue.  
Figure~\ref{f.sws} plots the spectra in the same colors.  In the
published paper, this figure will be Figure~19.
\label{f.aavso}}
\end{figure}

\begin{figure}[!ht] % Fig. 27
\includegraphics[width=240pt]{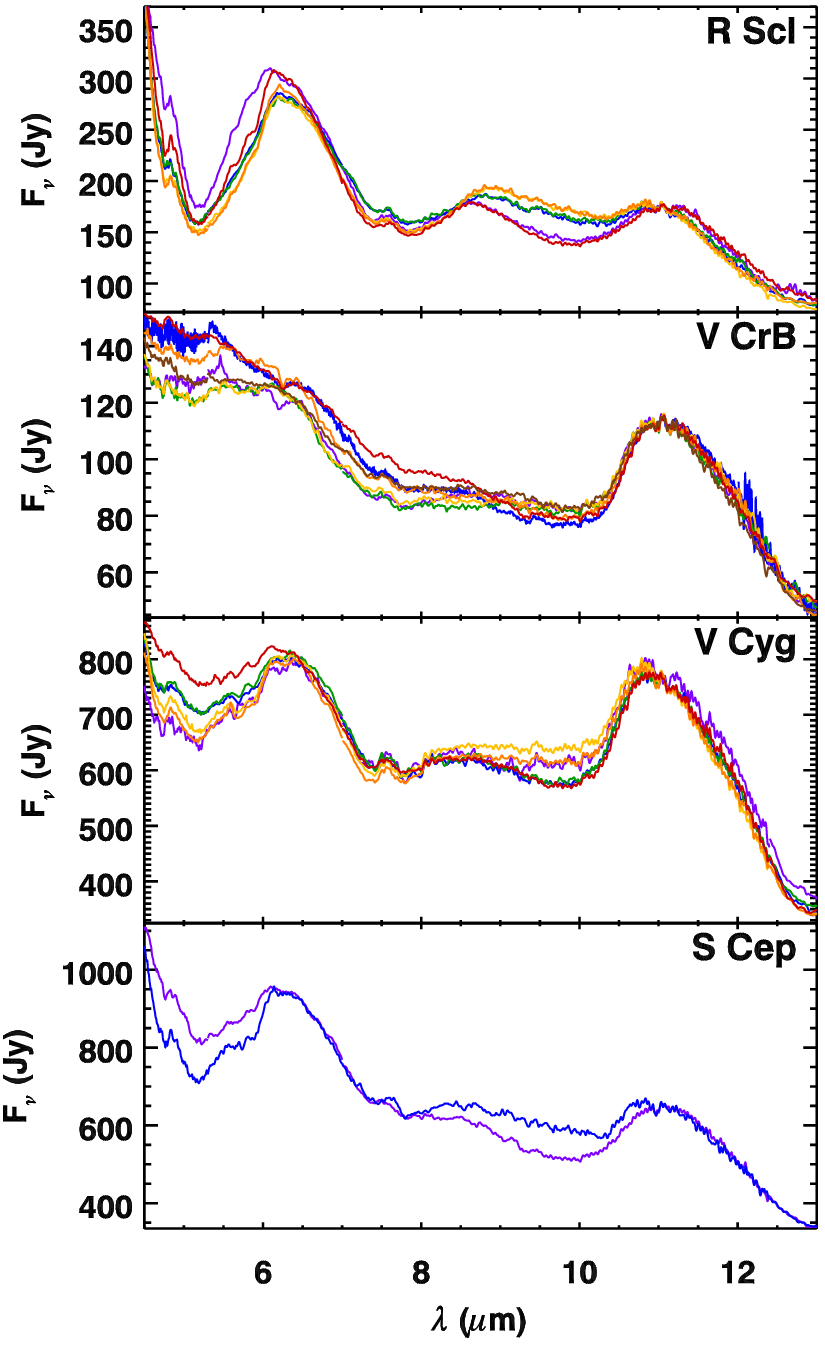}
\caption{Spectra from each SWS observation for the four Galactic 
carbon stars chosen to compare the behavior of the 10~\mum\ 
absorption band.  The spectra are plotted with the same colors as 
Figure~\ref{f.aavso}.  All spectra are normalized to the first
spectrum at 11.0--11.2~\mum.  In the published paper, this figure will 
be Figure~20.
\label{f.sws}}
\end{figure}

The SWS on ISO obtained reliable infrared spectra of 42 Galactic carbon
stars \citep{lei08, slo16}.  Several of these stars were observed
multiple times, and we examined the individual spectra of all of them
for variations in the apparent 10~\mum\ absorption band, using the 
spectra processed by \cite{slo03}.  Four sources showed this apparent 
absorption, and it varied in each of them.  

Table~\ref{t.galc} provides basic information about the four Galactic
carbon stars, including their [6.4]$-$[9.3] colors, which range from 
$\sim$0.3 to $\sim$0.5.  That interval includes the MRS targets 
J051803 and J053441.  R~Scl is an SRb variable, and the other three 
are Miras.  The pulsation periods range from $\sim$360 to 480 days, 
again similar to J051803 and J053441.

Table~\ref{t.sws} gives the details of each SWS observation 
considered.  Two observations of R~Scl and V~CrB occurred on the same 
day; we treated these independently, but they corroborate each other 
well.

To check the spectral variations with the pulsation phase, we 
analyzed visual photometry of the four Galactic carbon stars from 
the American Association of Variable Star Observers
\citep{klo25}.  We excluded upper limits and data described 
as ``discrepant.''  Figure~\ref{f.aavso} plots the AAVSO light curves, 
and it also shows when the SWS data were obtained.  For each SWS 
epoch, we estimated a visual magnitude by averaging the AAVSO 
data within an interval of $\pm$20 days.  While S~Cep has only two 
epochs, compared to six to seven  epochs each for the other three 
stars, they fall conveniently close to a visual maximum and minimum.

Figure~\ref{f.sws} plots the individual spectral observations from
the SWS.  The first and last observations of R~Scl, both taken at
or close to maximum, show stronger apparent absorption at 10~\mum\ 
than the four observations taken as the star approached minimum.  
The correlation of the strength of the band with phase is strong,
with the two observations closest to minimum showing the weakest
apparent absorption.  S~Cep shows the same dependency of apparent
10~\mum\ absorption with phase, with the caveat that we only have
two epochs.

If we had the same two observations of V~Cyg, we could expect to draw 
a similar conclusion, with the last epoch coming at maximum and 
showing deep apparent 10~\mum\ absorption, and the fourth epoch 
(gold) coming at minimum and showing the weakest apparent absorption.  
The remaining epochs, however, do not follow such a consistent 
pattern, with the second and third epochs (blue and green) obtained as 
the star approached minimum and showing almost as much apparent 
absorption as when the star was at maximum.  V~CrB also shows complex 
behavior.  While the epochs closest to minimum are among the spectra 
with the weakest apparent absorption (green, gold, and brown), and the 
epoch closest to maximum among the strongest (red), the epochs at 
intermediate phases do not appear to follow a coherent pattern.

We conclude that the SWS data confirm the presence of a varying
apparent absorption component at 10~\mum\ in the spectra of carbon stars 
observed with the MRS, and that the trend is generally for stronger
apparent absorption when the central star is at its maximum in the 
optical.

\end{document}